# Economic Complexity Unfolded: Interpretable Model for the Productive Structure of Economies

Zoran Utkovski[1,†], Melanie F. Pradier[2,†], Viktor Stojkoski[3], Fernando Perez-Cruz[4], Ljupco Kocarev[3,*]

**1** Fraunhofer Heinrich Hertz Institute, Berlin, Germany
**2** School of Engineering and Applied Sciences, Harvard University, Cambridge, MA, United States
**3** Research Center for Computer Science and Information Technologies, Macedonian Academy of Sciences and Arts, Skopje, Macedonia
**4** Swiss Data Science Institute (ETHZ/EPFL), Zurich, Switzerland

† Both authors contributed equally.
* lkocarev@manu.edu.mk

## Abstract

Economic complexity reflects the amount of knowledge that is embedded in the productive structure of an economy. It resides on the premise of hidden capabilities - fundamental endowments underlying the productive structure. In general, measuring the capabilities behind economic complexity directly is difficult, and indirect measures have been suggested which exploit the fact that the presence of the capabilities is expressed in a country's mix of products. We complement these studies by introducing a probabilistic framework which leverages Bayesian non-parametric techniques to extract the dominant features behind the comparative advantage in exported products. Based on economic evidence and trade data, we place a restricted Indian Buffet Process on the distribution of countries' capability endowment, appealing to a culinary metaphor to model the process of capability acquisition. The approach comes with a unique level of interpretability, as it produces a concise and economically plausible description of the instantiated capabilities.

## 1 Introduction

According to Lall [1, 2], each country has to find its own path towards development, focusing on its *learning system* in order to add *capabilities* to the ones it already owns. This line of reasoning, which Lall calls the "capabilities approach", has been further developed in the seminal works of Hidalgo and Hausmann [3–9], as well as in [10–12].

These fundamental endowments describing the productive structure of an economy are at the roots of the *economic complexity* theory, which leverages tools from network science and econometrics to reflect the amount of knowledge that is embedded in the productive structure of an economy (please see [3–6] and references therein for a detailed overview on the topic). According to the economic complexity approach, the complexity of the productive structure (more precisely the export structure) of countries is measured by using the information contained in the bipartite country-product network which links products and countries according to Balassa's Revealed Comparative Advantage (RCA) [13]. By using the information contained in the country-product matrix, a method has been proposed in [4] to derive *complexity* of countries and products as the components of the fixed point solution of an iterative linear map. As a result, countries in the international market are ranked and the difference in their competitiveness is measured based on their complexity score. With the intention to further reflect the ideas underlying the arguments of a capability driven economic competitiveness, the authors in [14, 15] propose a nonlinear relationship between the complexity of products and the fitness of countries. Both approaches have been shown to be economically-grounded and to be effective in ranking countries and products by their importance in the network [16]. Integration of services in the concept of economic complexity has been discussed in [17]. Besides highlighting the relationship between a country's productive structure and its economic growth, economic complexity essentially introduces non-monetary and non-income-based measures which uncover the countries' hidden potential for development and growth. As such, it sheds new light on the ongoing debate in the scientific community about the role of GDP as a measure for "economic success" [18, 19].

Recent work on productive structures has highlighted that the complexity and diversity of products a country exports are also a good proxy of the knowledge and know-how available in an economy that is not captured by aggregate measures of human capital [9]—such as the years of schooling or the percentage of the population with tertiary education. Moreover, productive structures can also be understood as a proxy of an economy's level of social capital and the health of its institutions, since the ability of a country to produce sophisticated products also critically depends on the ability of people to form social and professional networks [9]. Some of these issues are also addressed in the works of economic geographers which study path dependencies in the diversification processes of countries and regions [20–23]. Among others, in Refs. [24–27], measures of technological relatedness between patent classes were used to show that countries and cities develop new technologies related to existing local technologies. The distinctiveness of regional trajectories in economic geography is also closely related to the complexity of knowledge [28] and complexity of technology [29, 30].

Economic complexity has recently been also linked to income inequality [31], where a strong, robust, and stable correlation between a country's level of economic complexity (as proxied by the Economic Complexity Index [6]) and its level of income inequality has been established. The main argument goes along the lines that complex products tend to be produced by relatively few knowledge intense countries, and hence, can support higher wages for the workers employed in these industries (see also [6, 9]).

Despite the plethora of approaches, the general consensus in the literature on international and regional economic development is that economic complexity is reflected in a wide range of capabilities. The shared premise is that, instead of emerging randomly, new economic activities and knowledge build on and combine existing local capabilities, resulting in distinctive technological and industrial profiles of countries and regions.

## 1.1 Our Contribution

While there is a general consensus for a capabilities-driven productive structure, measuring the factors behind economic/technological/knowledge complexity directly is difficult, and indirect measures have been suggested which leverage the fact that the presence of these factors is expressed in a country's mix of products [3, 4, 6–12, 14, 15]. In the absence of a satisfying modeling framework, in the particular example of international trade, most of the research efforts have so far concentrated on the exported products by each country as the main proxy to infer the endowment of capabilities, i.e. the level of complexity of a productive system. Exception can be found in the binomial model for capabilities introduced in [5], and the work in [32] which estimates latent factors of endowments and formalizes the concept of latent comparative advantage.

Against this background, we extend over these methods by interpreting the productive structure from the perspective of probabilistic learning [33], where *latent features* are introduced to capture dependencies between exported products. This is similar in spirit to *Latent Semantic Analysis* in natural language processing, where documents are related to the words they contain via a set of topics [34]. In this same sense, the extracted features may be associated with the aforementioned capabilities, and may thus be understood as *factors* underlying the competitive advantage of countries in exported products.

However, in contrast to traditional factor analysis, the approach comes with the unique flavor of balancing predictive accuracy with *interpretability*, the current "holy grail" in high-dimensional data exploration. This interpretation is key for the subsequent exploitation phase which can be put forward as a discriminative model that provides highly accurate predictions. To assure interpretability and to simultaneously accommodate for the observed properties of real trade data (namely, the sparse triangular structure of the country product matrix [4, 14]), we propose a *Bayesian non-parametric* (BNP) approach which naturally encompasses sparse feature analysis when the underlying latent dimension is unknown. BNP models have mainly been used for clustering [35] and sparse feature analysis [36], when the number of clusters or features is a priori unbounded and is also learned from data. Many research disciplines have benefited from BNP models, including psychiatry [37], social sciences [38, 39], cancer research [40], or sports [41].

We note that, although being data driven, the method incorporates economic evidence about the relations between the different degrees of diversification in the exports and the expected distribution of the capabilities across the countries. In particular, we place an Indian Buffet Process [36] on the distribution of countries' capability endowment, appealing to a culinary metaphor to model the process of capability acquisition. Such flexible prior allows capturing different kinds of realities, from a world in which countries with few skills focus on different types of products, to a world in which less-developed countries have a strong overlap in export portfolios. Second, we also rely on the restricted Indian buffet process formulation from [42], which allows for a general marginal distribution over the number of active features per country.

The model not only allows us to characterize economic complexity of countries and products, but also to *unfold* the productive structure by isolating the key features associated with the competitive advantage in each of the exported products. With the capability interpretation, all measures (such as similarity, complexity, etc.) defined in the country or product space now admit a natural representation in the *capability space*. Moreover, we are able to identify features exclusively associated with less ubiquitous, i.e., more complex products. We refer to Fig. 1 to motivate the subsequent discussion and to illustrate the (first order) interpretative power of the model. It depicts the tripartite *country-capability-product network*, where the capabilities (latent features) relating countries and products are learned from data.

Conceptually, the framework is related to the works in economic development theory which reside on the ideas of implicit capacities, such as [32]. In particular, the authors in [32] perform non-linear principal component analysis (exponential PCA) as a dimensionality reduction technique to extract the latent variables which best explain the export data. However, compared to these approaches, the deployment of the Bayesian non-parametric approach with a flexible prior comes with the unique advantage of combining both interpretability and accuracy. As illustrated in the "Model Properties" section and in S1 Supporting Information, the comparison with the other factorization techniques suggests that our approach significantly enhances interpretability of the latent factors in terms of both conciseness and precision of the clustering of products. Additionally, we show that our findings are quite robust and stable in the sense that the evaluation of the model across different product classifications and years does not produce any significant changes in the results.

The proposed model exhibits further favorable qualitative and quantitative properties. Specifically, it captures the empirical distribution of ubiquity and, importantly, diversity, when compared to other models which fail to capture adequately the tails of the empirical diversity distribution. An important feature of the model is that it is able to incorporate temporal dynamics and thus capture the transitions in the capability space. The temporal dynamics can be put in the context of the studies that examine the path-dependence diversification in countries and regions (see e.g. [3, 43]).

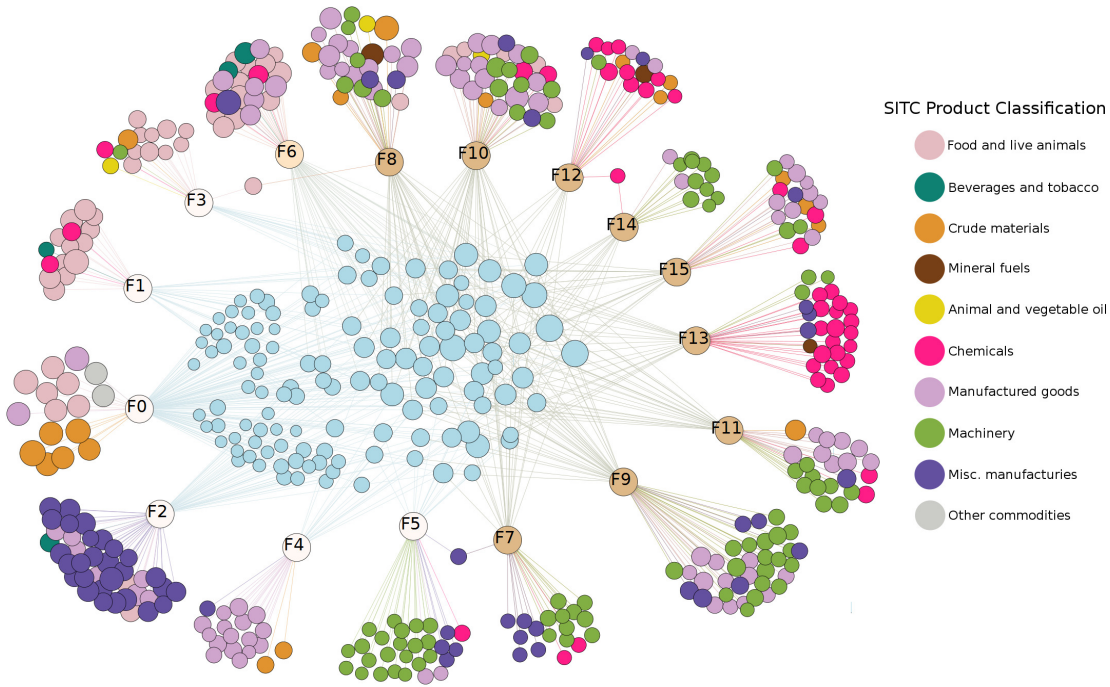

**Fig 1. The network of countries, capabilities and products**: A visualization of the tripartite network between countries, capabilities and products. In the middle of the network are the countries with node sizes proportional to their diversity. They are linked to the capabilities they have (the capability nodes are with uniform size). For visualization purposes, we link each capability $k$ to the products for which $B_{pk} \geq 0.3$, see Methods Section for further details. The size of the product nodes is proportional to their ubiquity and they are colored according to the one digit SITC classification. Products for which there is no $k$ such that $B_{pk} \geq 0.3$ are isolated, and thus are not shown in the Fig.

Consequently, our model indicates that export structures, being capability-dependent and thus difficult to change on short terms, have implications for growth and development. An interesting aspect of our model is the identification of latent features which are exclusively related to more diverse, i.e., developed and wealthier economies, as illustrated in the "Results" section.

## 2 Methods

We consider two publicly available trade datasets, the SITC and HS databases (see Section A of S1 Supporting Information for detailed description of the data). The data matrix is binary and represents the Revealed Comparative Advantage (RCA) of countries [13]. Basically, an entry $x_{cp}$ in the country-product matrix $\mathbf{X}$ equals one when country $c$ has a relative advantage at exporting product $p$, and zero otherwise. Even if the matrix is binary, we may use a Poisson likelihood because of the high degree of sparseness in data. Such approximation has already been adopted successfully in the case of recommendation systems [44].

The adjacency matrix $\mathbf{M}$ of the country-product network obtained from real trade data presents an approximately triangular sparse structure. This is in contrast to the block-diagonal structure as predicted by the Ricardian paradigm [45], which suggests that the wealthiest countries specialize in economic niches characterized by the production of only a few products with a high degree of specialization. The data thus suggests that countries have different diversity degrees in their export portfolios, and thus different trade strategies and skills. The main objective herein is to find an underlying representation that is easy to understand and able to capture this triangular structure in the input data.

## 2.1 Probabilistic Matrix Factorization

According to the proposed probabilistic framework, we assume that elements of $\mathbf{M}$ are realizations of random variables, whose dependencies are "captured" by *latent features*, i.e., capabilities which "dictate" relations between products. The elements of $\mathbf{M}$ are distributed according to a certain probability distribution $f$:

$$\mathbf{M} \sim f(\mathbf{ZB}), \tag{1}$$

where the $C \times K$ per-country (i.e., country-capability) matrix $\mathbf{Z}$ captures feature activation patterns, and the $K \times P$ per-product (i.e., capability-product) matrix $\mathbf{B}$ represents the effect of each latent feature on every product. For instance, if feature $k$ is active for a certain country, all products having high values in vector $\mathbf{B}_{k\bullet}$ will be more likely to be exported by that country. Under a slight abuse of notation, (1) might also refer to point-wise probability distributions, e.g., $M_{cp} = f(\mathbf{Z}_c \mathbf{B}_p)$. The approach may be interpreted as a probabilistic extension of *non-negative matrix factorization* where the number of latent features is not fixed a priori, both matrices are sparse, and soft-constraints on the expected latent sparsity structure are imposed through the prior.

## 2.2 Capability Endowment: A Culinary Metaphor

In order to model the capability endowment of countries, we place a modified Indian Buffet Process (IBP) prior over the per-country matrix $\mathbf{Z}$ [42, 46]. The standard IBP may be illustrated using a culinary metaphor which gives the name to the process [36]. Imagine an Indian restaurant whose buffet consists of infinitely many dishes arranged in a line. $C$ customers enter the restaurant sequentially. The first customer takes a serving from each dish, stopping after a Poisson $(\alpha)$ number of dishes, as his plate becomes full. The $c$-th customer moves along the buffet and samples dishes in proportion to their popularity, serving himself with probability $\frac{m_k}{c}$, where $m_k$ is the number of previous customers who tried dish $k$. Having reached the end of all previously sampled dishes, the $c$-th customer then tries Poisson $(\frac{\alpha}{c})$ new dishes. A matrix $\mathbf{Z}$ results from this experiment, such that $z_{ck} = 1$ when customer $c$ tries dish $k$. More generally, we say that a feature $k$ is active for sample $c$ if $z_{ck} = 1$. This process is denoted by $\mathbf{Z} \sim \text{IBP}(\alpha)$, where $\alpha$ is the mass parameter controlling the a priori activation probability of new features.

While an IBP may be appropriate to deal with the a priori unknown number of capabilities, the assumptions underlying the standard IBP are not flexible enough in this case. In fact, economic data suggests different capability distribution across countries, e.g., developed (diversified) countries should in general exhibit a higher number of capabilities. This is different to the standard IBP which implicitly assumes the same distribution Poisson($\alpha$) for the number of capabilities per country. We address this limitation via the Restricted IBP (R-IBP) formulation from [42]. The R-IBP allows for a general marginal distribution $g$ over the number of latent variables (capabilities) per sample (country). In particular, we choose $g$ to be a negative binomial in order to account for the over-dispersion of the number of capabilities per country. We further restrict the IBP by using a non-sparse bias term denoted as F0 which is active for all countries. Such term allows for the other latent features to be sparser and more interpretable [37, 41]. Finally, we combine the R-IBP with the three-parameter formulation from [46] that allows for different sharing degrees of capabilities across countries (potentially yielding a power-law distribution of the number of hidden features), increasing *de facto* the potential number of hidden capabilities.

### 2.3  Infinite Doubly-sparse Poisson Factorization

In addition to the prior for the country-capability matrix $\mathbf{Z}$, we specify a Gamma prior with shape parameter $\alpha_B$ and mean parameter $\mu_B$ for each element of the capability-product matrix $\mathbf{B}$. We enforce sparsity in the per-product matrix $\mathbf{B}$ by choosing $\alpha_B$ much smaller than one. Our model, that we refer to as Sparse Three-parameter Restricted-IBP (S3R-IBP), reads

$$M_{cp} \sim \text{Poisson}(\mathbf{Z}_{c\bullet}\mathbf{B}_{\bullet p});$$
$$B_{kp} \sim \text{Gamma}(\alpha_B, \frac{\mu_B}{\alpha_B});$$
$$\mathbf{Z} \sim \text{3R-IBP}(\alpha, \delta, \sigma, g), \qquad (2)$$

where $\mathbf{Z}_{c\bullet}$ denotes the $c$-th row of $\mathbf{Z}$, and $\mathbf{B}_{\bullet p}$ denotes the $p$-th column of $\mathbf{B}$. This model is fully specified by the a priori distribution $g$ over the number of active features per sample, together with three hyperparameters: i) $\alpha$, which is the same mass parameter from the standard IBP controlling the *a priori* total number of non-zero entries in matrix $\mathbf{Z}$; ii) $\sigma \in [0, 1)$ is the stability exponent which controls the power-law behavior of the model; iii) $\delta > -\sigma$ is the concentration parameter that affects the sharing degree of capabilities across countries. More details about the methodology can be found in Sections B-C of S1 Supporting Information and [47].

Altogether, our model has been designed to find highly-specific and easy-to-interpret latent features involving only a few products and are active for a small number of countries, in consistency with the economic literature [5]. Since exact computation of the posterior distribution for the latent variables is intractable, inference is performed using a Markov Chain Monte Carlo procedure, as described in Section D of S1 Supporting Information.

## 3  Results

In the following we present a qualitative evaluation of the model. In particular, we introduce the capability space and discuss the interpretative power of the model. We also address correlations in the capability space and their implications. Finally, we introduce temporal dynamics, with the idea to capture the dynamics in the acquisition of capabilities and thereby associated products over time. Quantitative evaluation is provided in the subsequent Section (Model Properties).

## 3.1 The Capability Space

By adopting the probabilistic learning framework, we are able to *unfold* the productive structure by isolating the key features associated with the competitive advantage in each of the exported products. As a result of the decomposition of the bipartite country-product network, complexity-related measures (such as country/product complexity, product similarity etc.) now admit a natural representation in the *capability space*.

In the first step, through the extraction of capabilities we are able to separate highly-diversified from less-diversified countries. Countries with low diversity are associated with less capabilities, and these capabilities are linked to more ubiquitous products. This observation is in line with [4, 14], as it suggests that capabilities have different degrees of "complexity", which plays an important role in the production (export) of products. Moreover, the model produces homogeneous and concise clustering of products, i.e., products of similar ubiquity and SITC 1 digit classification are usually associated to the same capability

This effect is also captured in Fig. 1, where we connect countries to capabilities whenever $Z_{ck} = 1$, and capabilities to products whenever $B_{kp} \geq 0.3$. The size of the country nodes is proportional to their diversity $d_c$, defined as the number of products in which the country has comparative advantage. Similarly, the product node size is proportional to their ubiquity $u_p$, which corresponds to the number of countries exporting that product having comparative advantage (see Section A of S1 Supporting Information). This implies that the differentiation of capabilities takes place according to both the level of diversification of countries and the elements required in the production of products.

Table 1 captures the interpretable power of the model, by listing the capabilities found by the S3R-IBP model in 2010 (SITC classification of products). For each of the instantiated capabilities, we report the averaged number of countries endowed with the respective capability, as well as the top-5 products associated with it. Also, for each capability in the list, we single out a *representative country*, defined as the one with the smallest number of active capabilities among all countries possessing the capability in question. For each representative country, we also report the average number of active capabilities.

**Table 1. Complete list of capabilities found by the S3R-IBP model in 2010 through the SITC classification.**

| Id | $\bar{m}_k$ | Top-5 products with sorted highest weights ($B_{kp}$) associated | Repr. countries ($\bar{J}_c$) |
|---|---|---|---|
| F0 | 126 | Non-Coniferous Worked Wood (0.40), Bran and Other Cereals Residues (0.39), Miscellaneous Non-Iron Waste (0.38), Unwrought Lead (0.38), Bones, Ivory and Horns (0.37) | – |
| F1 | 38.67 | Vegetables (0.60), Fruit or Vegetable Juices (0.54), Miscellaneous Fruit (0.50), Frozen Vegetables (0.48), Apples (0.47) | Peru (2.00) |
| F2 | 46.11 | Synthetic Knitted Undergarments (0.76), Miscellaneous Feminine Outerwear (0.74), Miscellaneous Knitted Outerwear (0.73), Men's Shirts (0.70), Blouses (0.67) | Sri Lanka (2.00) |
| F3 | 18.27 | Miscellaneous Animal Oils (0.78), Bovine and Equine Entrails (0.72), Bovine meat (0.68), Preserved Milk (0.63), Equine (0.62) | Paraguay (2.00) |
| F4 | 21.39 | Synthetic Woven Fabrics (0.74), Non-retail Synthetic Yarn (0.60), Woven Fabric of less than 85% Discontinuous Synthetic Fibres (0.60), Woven Fabrics of More Than 85% Discontinuous Synthetic Fiber (0.58), Yarn of Less Than 85% Synthetic Fibers (0.53) | United Arab Emirates (2.82) |
| F5 | 16.53 | Miscellaneous Electrical Machinery (0.76), Vehicles Stereos (0.72), Miscellaneous Data Processing Equipment (0.64), Video and Sound Recorders (0.57), Calculating Machines (0.55) | Malaysia (3.00) |
| F6 | 45.93 | Baked Goods (0.67), Metal Containers (0.62), Miscellaneous Edibles (0.59), Miscellaneous Articles of Paper (0.59), Miscellaneous Organic Surfactants (0.58) | Costa Rica (2.06) |
| F7 | 21.95 | Measuring Controlling Instruments (0.61), Mathematical Calculation Instruments (0.59), Miscellaneous Electrical Instruments (0.57), Miscellaneous Heating and Cooling Equipment (0.51), Parts of Office Machines (0.49) | Malaysia (3.00) |
| F8 | 33.23 | Miscellaneous Articles of Iron (0.65), Carpentry Wood (0.61), Miscellaneous Manufactured Wood Articles (0.60), Sawn Wood Less Than 5mm Thick (0.56), Electric Current (0.51) | Russia (2.93) |
| F9 | 32.12 | Miscellaneous Rotating Electric Plant Parts (0.66), Control Instruments of Gas or Liquid (0.58), Valves (0.57), Miscellaneous Rubber (0.56), Miscellaneous Articles of Plastic (0.55) | Philippines (4.01) |
| F10 | 33.00 | Improved Wood (0.71), Mineral Wool (0.62), Central Heating Equipment (0.62), Aluminium Structures (0.62), Harvesting Machines (0.60) | Belarus (4.20) |
| F11 | 31.14 | Vehicles Parts and Accessories (0.59), Cars (0.58), Iron Wire (0.53), Trucks and Vans (0.53), Air Pumps and Compressors (0.50) | Belarus (4.20) |
| F12 | 11.04 | Synthetic Rubber (0.87), Acrylic Polymers (0.85), Silicones (0.76), Miscellaneous Polymerization Products (0.71), Tinned Sheets (0.65) | North Korea (3.99) |
| F13 | 18.67 | Aldehyde, Ketone and Quinone-Function Compounds (0.68), Glycosides and Vaccines (0.67), Medicaments (0.65), Inorganic Esters (0.64), Cyclic Alcohols (0.62) | Ireland (4.34) |
| F14 | 14.87 | Parts of Metalworking Machine Tools (0.74), Interchangeable Tool Parts (0.72), Polishing Stones (0.69), Tool Holders (0.66), Miscellaneous Metalworking Machine-Tools (0.54) | Israel (5.97) |
| F15 | 23.29 | Miscellaneous Pumps (0.51), Ash and Residues (0.45), Chemical Wood Pulp of sulphite (0.44), Rolls of Paper (0.43), Worked Nickel (0.43) | Russia (2.93) |

Notes: From left to right, $\bar{m}_k$ is the averaged number of countries having latent feature $k$ active, we list the top-5 products with highest weights $B_{kp}$; a *representative country* is the country that has the least number of capabilities among those possessing feature $k$. $\bar{J}_c$ is the averaged number of active features for each representative country $c$.

Fig. 2 presents a graph-based description of the capability space, which aims at capturing correlations between the capabilities. Two capabilities $k$ and $j$ are highly correlated (share an edge) if they co-occur frequently across countries, which can be measured for example by using the Jaccard similarity index between column vectors $\mathbf{Z}_{\bullet k}$ and $\mathbf{Z}_{\bullet j}$. In fact, capability co-activation patterns may be potentially useful to define policy recommendations in a subsequent exploitation phase: as capabilities are related to each other, having one capability active increases the likelihood of having other capabilities active. For example, for a country possessing capability $k$, but not $j$ (in the case when $k$ and $j$ are significantly correlated), the acquisition of $j$ may come at a relatively small cost, thus justifying policy incentives in this direction.

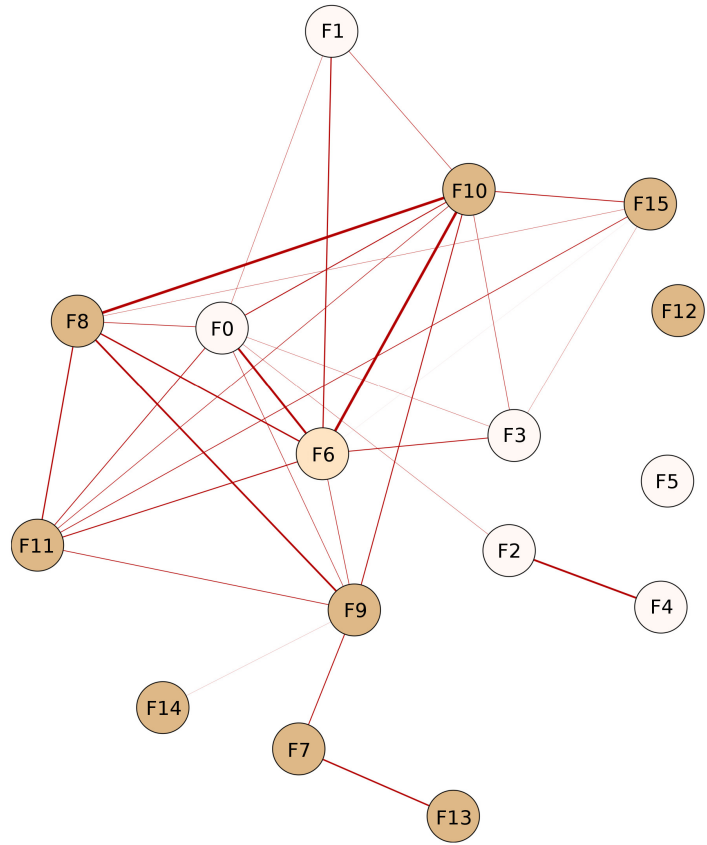

**Fig 2. Correlations in the capability space.** Nodes correspond to inferred capabilities. The coloring is according to the meta-capability grouping (subsection "The Meta-Capability Space"). Edge width and intensity are proportional to the correlation strength. For better visibility, we only depict edges with correlation higher than 0.4.

In the following we summarize some general observations for the inferred correlations between capabilities in our model: 1) Capabilities F5 and F12 associated with electronics and chemicals respectively, are not highly correlated with the rest; 2) F4 (synthetic fabrics and fibers) and F2 (clothing items) are highly correlated, which is logical as both relate to the clothing industry, and are almost isolated from the rest of the graph; 3) F1-F5 are representative for developing countries, as explained in the next subsection; 4) F14 (metalworking machine-tools) is loosely related to F9 (misc. rotating electric plant parts, control instruments, but also misc. rubber and plastic products) only; 5) F6 is particularly interesting, as it is associated with heterogeneous products such as baked goods and misc. edibles, but also metal containers, misc. articles of paper, and misc. organic surfactants. F6 has the role of a "capability hub" towards more advanced (i.e. less ubiquitous) capabilities such as F11 (vehicles), F9, F8 (iron and wood articles) and F10; 6) F3 and F1 (roughly farming and agriculture) seem to be the starting point of developing countries to acquire some more advanced skills, in particular captured by F6 and F10; 7) F13 (bio-chemical products such as medicines, vaccines and carbonyl compounds) seems to be the most difficult capability to be acquired, as it is disconnected from the "core" of the network and is related only to F7 (specialized electronics, heating and cooling machines, and parts of office machines), itself being also advanced and relatively isolated from the rest of the network.

## 3.2 The Meta-Capability Space

In order to further analyze the relationships between the inferred capabilities, we again apply our S3R-IBP procedure over the inferred capability activation matrix **Z** as input data. Such a deep structure, i.e. using two-layer IBP, has already been explored in [48]. As before, we use a bias term, denoted by M-F0, which is active for all countries. In addition to M-F0, our approach extracts another meta-feature (meta-capability), M-F1, which is active for 46.19 countries on average. The list of countries is provided in Table 2. Both M-F0 and M-F1 assign different weights to each capability from the first layer. The countries with an active M-F1 are those that have more capabilities in the original model and, in general, are more diversified. Hence, we interpret M-F1 as the meta-feature that separates countries based on their level of diversification.

Table 2. Meta-features activity pattern

| MF-0 | MF-1 | List of Countries having those activation patterns for the meta-features |
|:---:|:---:|:---:|
| 1 | 0 | Pakistan, Syria, Chile, Kyrgyzstan, Zimbabwe, Albania, Tanzania, Bahrain, Laos, Botswana, Bolivia, Bangladesh, Kazakhstan, Senegal, Cuba, Zambia, Namibia, Oman, Turkmenistan, Mongolia, Ethiopia, Mozambique, Iran, Ghana, Cote d'Ivoire, Papua New Guinea, Saudi Arabia, Yemen, Sudan, Trinidad and Tobago, Cameroon, Mauritania, Venezuela, Guinea, Azerbaijan, Algeria, Republic of the Congo, Kuwait, Nigeria, Qatar, Gabon, Libya, Iraq, Angola |
| 1 | 1 | Germany, Italy, United States, Japan, France, China, Austria, Czech Republic, Spain, United Kingdom, Belgium, Sweden, Netherlands, Switzerland, Poland, Denmark, Portugal, Hong Kong, India, Slovenia, Finland, Hungary, Thailand, Israel, Turkey, South Korea, Slovakia, Bulgaria, Romania, Croatia, Estonia, Serbia, Canada, Lithuania, Singapore, Mexico, Panama, Ukraine, Latvia, Malaysia, Brazil, Indonesia, Greece, Bosnia and Herzegovina, Tunisia, Lebanon, Ireland, Vietnam, Philippines, Argentina, Belarus, Egypt, South Africa, North Korea, New Zealand, Russia, Uruguay, El Salvador, United Arab Emirates, Norway, Morocco, Sri Lanka, Moldova, Macedonia, Jordan, Colombia, Australia, Kenya, Mauritius, Peru, Guatemala, Uzbekistan, Dominican Republic, Paraguay, Madagascar, Costa Rica, Honduras, Georgia, Ecuador, Nicaragua, Cambodia, Burma |

These two meta-features furthermore divide the capabilities learned in the original model into three disjoint sets. The first set contains the latent features whose weight is either zero or insignificant for M-F1, F0 to F5 (highlighted in white in Fig. 1). These are the features that define countries which deal with goods whose production mostly relies on the presence of physical factors (i.e. natural resources). Hence, it makes sense that more diversified countries do not have stronger weights than less diversified countries for these features. While diversified countries might have a non-zero RCA in the products associated with these features, they are not necessarily exploiting them better than the less diversified countries.

The second set consists of a single feature, F6 (highlighted in light yellow in Fig. 1),

which has a high weight in both M-F0 and M-F1. This feature is associated with
products exported by both groups of countries. However, the diversified countries do
trade these products more efficiently than the less diversified ones. We can interpret the
products associated with this feature as being on the capability frontier.

The last set includes the remaining features (highlighted in gold in Fig. 1), with
weights in M-F0 which are negligible compared to their weights in M-F1. These features
are associated with less ubiquitous products, for example chemicals and complex
machinery (see Table A in S1 Supporting Information), which are only exported by
diversified (i.e. economically complex) countries. To put the findings in geographic
context, in Fig. 3 we plot a heat map of the world according to the presence of features
associated with meta feature M-F1.

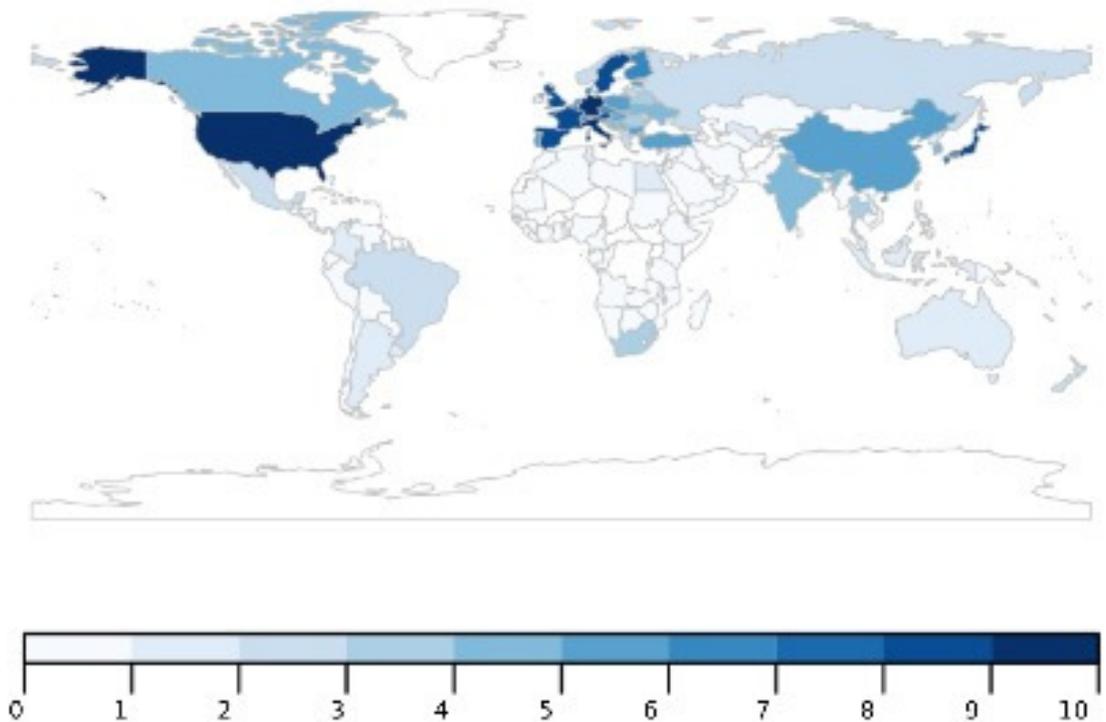

**Fig 3. Meta feature activation.** World heat map according to the presence of features associated
with meta feature M-F1. Darker shade indicates presence of more capabilities that are associated with
M-F1. Countries with the lightest blue shade only have M-F0 capabilities, whereas there is no data for
the countries in white shade. The map was generated in the software R (Available at
https://cran.r-project.org/) using the package "rworldmap" [49] and data from authors' own
calculations.

### 3.3 Countries in the Capability Space

Given the factorization (1), it is possible to define distances between countries based on
their capability endowment, i.e., we may define a similarity measure in the capability
space. Effectively, by comparing the latent representations of two countries in the
capability space, we can say in which way two countries are similar. In this sense, the
framework provides a complementary representation to the approaches which measure
similarity directly in the product space.

For illustration of the concept, we use a simple similarity measure based on the weighted Euclidean distance between the vectors representing the capability endowments of individual countries, where the weights are inversely proportional to the capabilities ubiquity. To be precise, weights are proportional to $(1 - \pi_k)$ where $\pi_k$ is the proportion of countries having capability $k$. Intuitively, two countries are more similar if they share a less ubiquitous capability. The mapping to the two-dimensional plane is performed via non-classical multidimensional scaling [50].

Fig. 4 depicts the computed distances between countries in 1995 and 2010. In both years we observe a core of closely spaced countries with similar underdeveloped productive structure, i.e. with minimal capability endowment. Moreover, in 1995 we notice a relatively strong polarization in the world due to large differences between the most and least diverse economies. Compared to 1995, the situation in 2010 is somewhat less polarized, as some of the less diverse countries gradually acquired capabilities over the years, increased their diversification and consequently narrowed the gap to the most diverse countries.

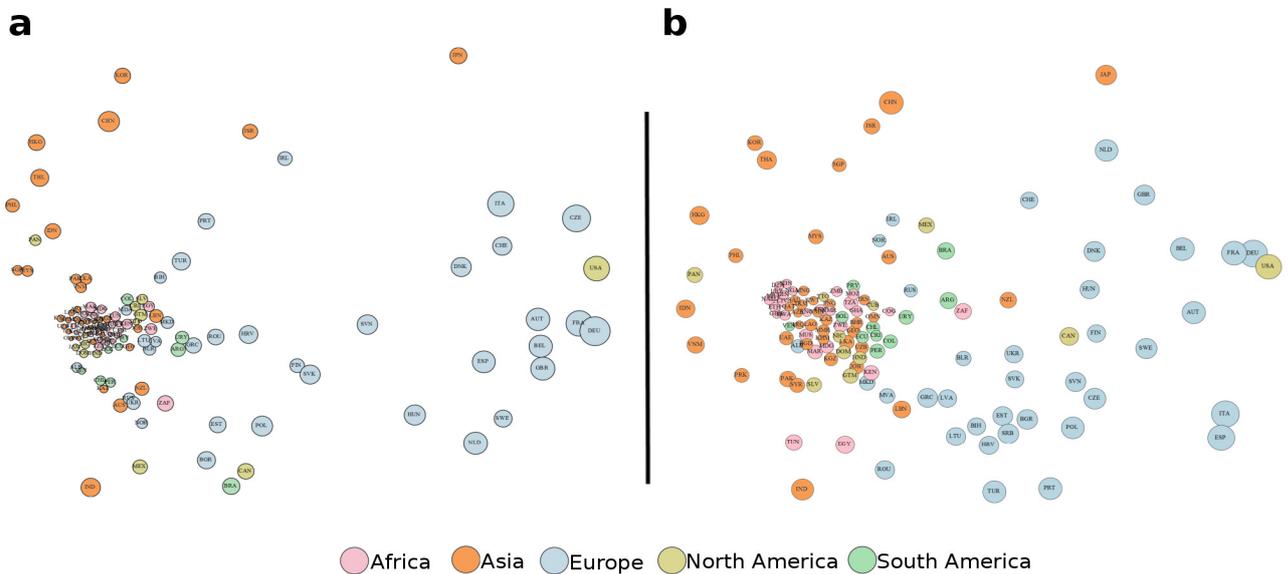

**Fig 4. Countries in the capability space.** The distance between two countries is an approximation of the Euclidean distance between their capability vectors estimated via Multidimensional Scaling. Node size is proportional to country's diversity, whereas node color is region-based. **(a)** Calculated with data for 1995. **(b)** Same as **(a)** only for 2010.

The results from Fig. 4 are in line with a growing body of empirical studies which suggest that knowledge diffuses among neighbors and/or that shared history influences the future development [51–53]. The figure also reveals interesting relations between the productive structures of individual countries within regions and across those having different economic policy ideologies. For instance, we can easily separate the Asian economies into two groups, one in the left periphery and another at the core of the Figure. The first group are countries which exhibited significant economic development over the examined years, whereas the countries in the other group stagnated. Typical countries from the first group are the "Asian Tigers" Hong Kong, Singapore and South Korea (note that Taiwan is also in the same group, but it does not satisfy our criteria for entering the dataset), placed in the top left corner of Fig. 4 **a** and Fig. 4 **b**. A distinct characteristic of these economies was the exploitation of previously unused industrial capabilities for developing information and communication based technologies (ICT) that contributed to their economic growth rates. Fig. 4 also presents interesting outliers, such as the Republic of Ireland, whose productive structure differs substantially from those of their geographic neighbors. Interestingly, Ireland was also dubbed as *Celtic Tiger* due to the economic characteristics and the implementation of policies similar to the Asian Tigers.

Fig. 5 depicts the positioning of the Tigers in the capability space, based on their inferred capabilities. All four countries share two capabilities, F0 (basic capability present in all countries), and F5, which is exactly associated to ICT intensive products, as shown in Fig. 1 and Table 1. This example suggests that our model can be effectively utilized as a discriminative framework to explain country economic dynamics based on neighborhood relationships in the capability space.

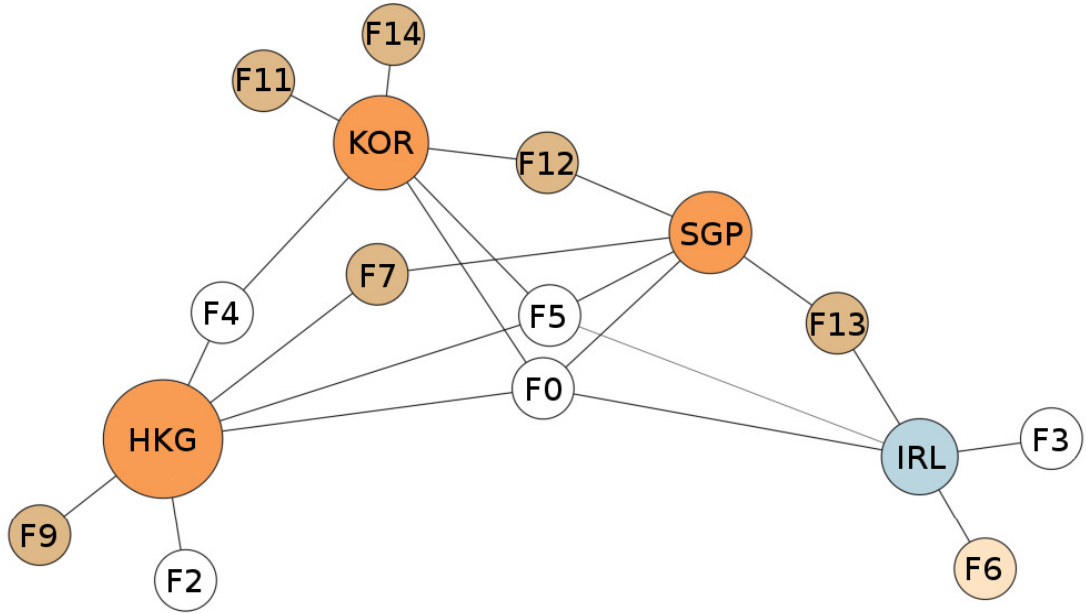

**Fig 5. The Tigers' capabilities.** A zoom in on the positioning of Hong Kong, Ireland, Singapore and South Korea in the capability space in 2010. We also show the capabilities and link the countries to the capabilities they have. For simplicity, we exclude the capabilities that are absent in all four countries.

In general, it would be relevant to interpret these findings from the perspective of evolutionary economic geography, see e.g. [21, 43, 54].

### 3.4 Products in the Capability Space

Hidalgo et al. [6] propose a similarity measure between goods by looking at the probability that they are co-exported. To quantify this similarity it is assumed that if two goods share most of the requisite capabilities (where the reference to the capabilities is implicit), the countries that export one will also export the other. By analogy, it is expected that goods that do not share many capabilities are less likely to be co-exported. The proximity measure intents to infer the similarity between the capabilities required by a pair of goods by capturing the conditional probability that a country that exports product $p$ will also export product $p'$. In the absence of explicit reference to the underlying capabilities, proximity is inferred directly from the entries in

the $M$ matrix as

$$\phi_{p,p'} = \frac{\sum_c M_{cp} M_{cp'}}{\max(k_{p,0}, k_{p',0})} \tag{3}$$

where $k_{p,0}$ is the ubiquity of product $p$. Since conditional probabilities are not symmetric, in (3) the minimum of the probability (which is inversely related to the maximum of the ubiquity) of exporting product $p$, given $p'$ and the reverse, is considered.

In our framework, which provides a full probabilistic description of the trade flows with explicitly instantiated capabilities, product similarity may naturally be defined in the capability (i.e. the feature space). Here, a starting point is the representation of the products by the column vectors of the matrix **B** in the decomposition (2), which interprets products as points in the capability space. Based on this representation, various "similarity" measures between products may be introduced. In the first approximation, the framework offers an efficient clustering of products by adding a threshold in the vector representation. As depicted in Fig. 1, this rather simple methodology produces a compact and consistent representation of products. While products of similar ubiquity and SITC one digit classification are often associated to the same capability, our approach is able to furthermore differentiate products inside this broad classification. For instance, capabilities F1 and F3 separate vegetable and fruits from animal products, which appear together in the SITC 0 as "Food and live animals group". As such, our model may provide alternatives to the SITC classification of products, based on their positioning in the capability space.

A more refined approach relying on a high-dimensional embedding of the per-product matrix **B** could bring additional insights on the relations between different products.

## 3.5 Temporal Dynamics

A large volume of studies provides strong evidence supporting the notion that diversification in countries and regions is path dependent. The authors in [3, 43] showed how countries expand their mix of exports around the products in which they already established a comparative advantage. Neffke et al. [21] used information on product portfolios to show that regions tend to diversify into new industries related to existing local industries. Refs. [24–27], among others, used measures of technological relatedness between patent classes to show that countries and cities develop new technologies related to existing local technologies.

We complement these studies by introducing a temporal dynamics in our probabilistic model. This is performed by applying the S3R-IBP model to the aggregated SITC database between 1964 and 2010. We assume a shared set of latent features over the years, and independent feature activations for each year, e.g., USA in 1965 is counted as a different country from USA in 1986. To speed up mixing, we initialize the features to the values obtained from the analysis of 2010 data, listed in Table 1, and learn the feature activation values for all years.

Fig. 6 shows the feature activation dynamics for three particular examples: Chile, Egypt and Indonesia. All these countries were able to increase their number of active features over the years, although their dynamics can be attributed to different economic factors. The latent features may be used to infer information for the internal situation of the countries' economic system at different time instants. In particular, the growth of Chile can be attributed to the activation of features associated to natural resources, whereas the growth of Indonesia is due to acquisition of features related to clothing and electronic products. Interestingly, Chile is a well known natural resource producer [6], while the start of Indonesia's growth coincides with the period when the country opened its economy and got an influx of foreign direct investments.

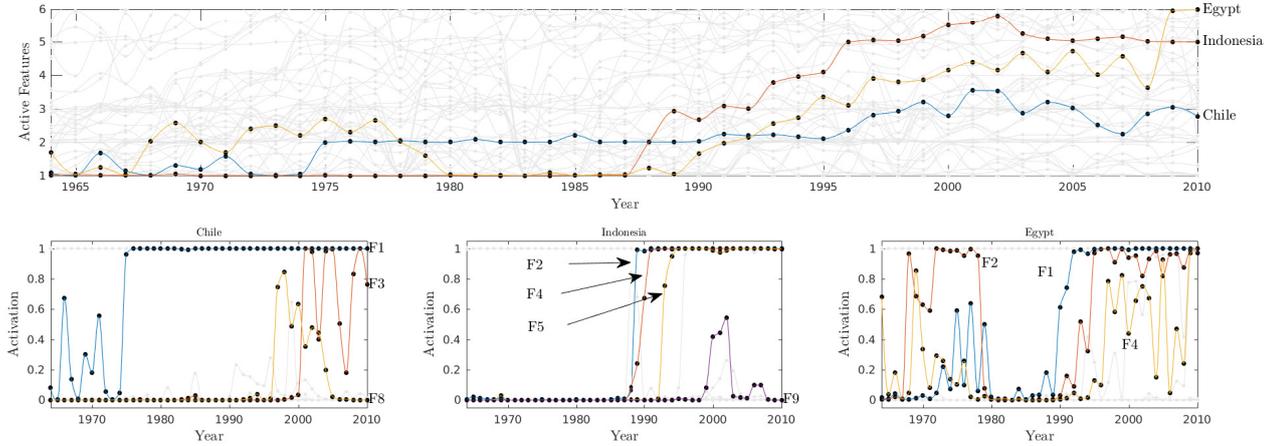

**Fig 6.** Dynamics for selected countries. Top row: number of active features per year for Chile (blue), Egypt (red) and Indonesia (green). Bottom row: activation of features for the same countries.

Finally, Egypt is a country that has very unique dynamics. There is a sudden fall in the activity of features F2 and F4 (both associated to clothes) and F1 (vegetables and fruits) at the end of the 1970s, after which the number of active features remained steady at its minimum, until the beginning of the 1990s. The steady state period corresponds to years of political and economic instability [55]. At the beginning of the second growth, which corresponds to a period of political reforms that made Egypt a more open economy, the country regained the activity in F1, F2 and F4 quickly, and even adding vehicles to its export basket (F11).

In addition to monitoring the temporal dynamics of each individual country, it is also interesting to study the feature transitions globally, as a simple transition model. This is possible given the discrete nature of the capability endowments, as features can be either active or inactive (binary vector). To better account for time, we retrained our model assuming a Markovian dependency across capability endowment vectors $\mathbf{Z}_{c\bullet}(t)$ for each country $c$ over the years. The Markovian assumption works as a smoothing factor of the country trajectories in the capability space. In fact, $\mathbf{Z}_{c\bullet}(t)$ can be interpreted as the latent state of country $i$ at time $t$, which indicates the set of features that are active for that country at that particular point in time. Let $G = \{M, E\}$ be a hidden network, where $M$ denotes all possible latent states (all possible values for vector $\mathbf{Z}_{c\bullet}(t)$ and $E$ refers to all directed edges connecting any two elements belonging to $M$ (full network). For each country $i$, the sequence $[\mathbf{Z}_{c\bullet}(1), \mathbf{Z}_{c\bullet}(2), \ldots, \mathbf{Z}_{c\bullet}(T)]$ corresponds to a path within such network. By monitoring which are the most common nodes and most visited edges in the network, we can gain new insights on the developing mechanisms of countries. An illustration of the concept is provided in Fig. 7 corresponding to the first steps in countries' development. Fig. 7 depicts a subset of the transition model induced by the temporal evolution of countries in the capability space. Node 1 corresponds to the state in which only the bias term is active (no other capabilities), where 16% of the countries are placed each year on average. From that state, the easiest path is to acquire G1 (coffee, sugar and wood plantations), after which countries might get G2 (basic manufacturing), G3 (clothing) or both.

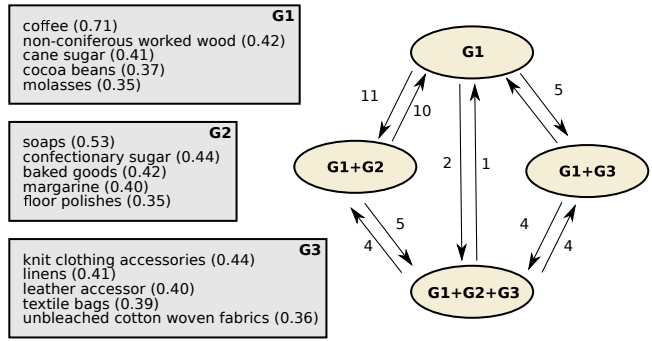

**Fig 7. Subset of the transition model**: Each node corresponds to the set of active features at this state; edges are weighted by the corresponding number of inferred transition in **Z**.

## 3.6 Possible Effects on Policy Recommendations

The product space analysis in [3] suggests that there is a tendency for countries to develop RCA close to products for which RCA was already developed. According to the terminology in [3], a product for which a country has developed RCA is called an occupied product (O), and unoccupied product (U) otherwise. By considering the transition in the product space which takes unoccupied products to occupied ones (U → O), the authors in [3] provide evidence that countries perform structural transformations by jumping from occupied products to nearby ones. The analysis involves the definition of a quantity referred to as *density*, which is defined as a weighted fraction of the space which is occupied from the point of view of a product in a particular country. Similarly, one can consider the probability for a product to develop given that the closest developed product is at a certain proximity, as defined by the authors in [6].

By employing our probabilistic approach, we connect to the observations in [3] and [6]. Specifically, we define a simple quantity that can be defined from the inferred country-capability-product decomposition. In particular, the estimated likelihood that country $c$ can export product $p$ given the country's capability vector $\mathbf{Z}_{c\bullet}$, can be used to define two sets of products. The first set includes those products for which the country has high likelihood according to our model, but no RCA has been developed yet. Conversely, the products for which that country has low likelihood but has already comparative advantage are those that are at risk to be lost in the future, as inferred from the capability decomposition. We note that this line of reasoning aligns with the concept of latent comparative advantage, which has been introduced in [32], and aims to assess how well countries are matching their potential implied by the latent endowment, as well as to identify products for which the latent advantage is not yet revealed (extensive margin).

To investigate these hypotheses, and to provide complementary insights to the works in [3], [6] and [32], we monitored the products with RCA=1 and highest probability of deactivation (transition O → U as described in [3]), as well as the products with RCA=0 and highest probability of activation (transition U → O as described in [3]) in 2010. We thereby estimated the percentage of those products that get activated in the next 4 years. This percentage ranges within $32 - 44\%$ (the variation depends on which threshold parameter is chosen to transform the soft-predictions of our model into binary RCA values), in contrast to a random activation of 7% (when choosing products at random that have RCA=0). Table 3 and Table 4 present the top-5 products with RCA=1 and highest probability of deactivation, as well as RCA=0 and highest probability of activation respectively, for the subset of countries studied in the "Temporal dynamics" subsection.

In each country, note that the exported products at risk are those for which the required capabilities are not present (in a probabilistic sense, and with respect to a threshold). For instance, in Indonesia and Egypt these are the exported chemicals (or products used in chemistry) whereas in Chile the products with the highest chance to be deactivated are machineries. On the other hand, the products that the countries are most likely to occupy in the future (i.e. to add to their export baskets) are related to the capabilities already acquired by the respective countries. Note that the most likely new products for Chile are miscellaneous manufactured goods, which are related to feature F8. This is a capability in which Chile had a high weight in the past (see Fig. 6), and thus, might be more likely to get it active in the future.

Table 3. Monitoring products at risk

| Countries | Exported products at risk (lowest weights) per country based on our model |
|---|---|
| Chile | Crude Natural Potassium Salts (0.03), Toys and Games (0.03), Nuclear Reactors (0.04), Metal Cutting Machines (0.04), Miscellaneous Metalworking Machine-Tools (0.04) |
| Egypt | Photographic Chemicals (0.03), Sulphonamides, Sultones and Sultams (0.03), Miscellaneous Indoors Sanitary Ware of Base Metal (0.04), Baby Carriages (0.04), Castor Oil Seeds (0.04) |
| Indonesia | Copolymers of Vinyl Chloride and Vinyl Acetate (0.02), Silicones (0.02), Steam Power Units (0.03), Natural Sodium Nitrate (0.03), Photographic Chemicals (0.04) |

Table 4. Incorporating new products in the export portfolio

| Countries | Promising products (highest weights) per country based on our model |
|---|---|
| Chile | Aluminium Structures (1.15), Cotton Yarn (0.91), Inorganic Chemical Products (0.82), Live Plants (0.80), Uninsulated Steel Wire (0.77) |
| Egypt | Bovine meat (1.01), Bonded Fiber Fabrics (0.90), Umbrellas and Canes (0.85), Fiberboard (0.84), Leather Accessories (0.83) |
| Indonesia | Valves (1.12), Unmilled Oats (0.92), Metal Cables (0.92), Metal Office Products (0.91), Acrylic Polymers (0.81) |

As a final note, we suggest that the temporal model and the finite state machine interpretation might provide a step towards the more fundamental understanding of the typical evolution paths of developing countries, ideally followed by possible policy recommendations in the light of these findings. The evaluation of the effects of specific policy recommendations, which may, for example, come in the form of subsidies or R&D incentives that the countries may provide to stimulate the production of specific products, is not covered by this work.

## 4 Model Properties

To describe the general properties of our model, in Fig. 8 **a-b** we depict the empirical and inferred country-product adjacency matrices (**M** matrix) for 2010. The countries are ordered according to their decreasing diversity $d_c$, whereas products are sorted according to their decreasing ubiquity $u_p$. It is evident that the model successfully reproduces the triangular sparse structure present in the empirical data matrix. Fig. 8 **c-d** depicts the inferred country-capability matrix **Z**, and capability-product matrix **B** for the same year. The country-capability matrix also exhibits a sparse triangular structure, suggesting that more diversified countries also have more capabilities. On the other hand, the weights in the capability-product matrix are block sparse with varying sparsity among the rows (capabilities), implying that there are certain capabilities which are associated with more products.

Furthermore, in order to evaluate the performance of our model, in Fig. 8 **a-f** we compare the inferred cumulative distribution of our S3R-IBP model with two other models. The first model used in the comparison is a simple binomial model described in [4], which assumes binary matrices **Z** and **B**, a finite number of capabilities $K$, and uniform activation probabilities $r$ and $q$ for all country-capability and capability-product combinations. The parameters in this baseline model are chosen to best fit a functional form connecting a country's diversification to the average ubiquity of its products [5]. The second model we compare with is the IBP with sparse Gamma prior on the features with a bias term (S-IBP).

We observe that all models adequately reproduce the distribution of ubiquity. However, there are large differences when comparing the models' ability to explain the distribution of diversity. In particular, the baseline model predicts an almost uniform distribution, whereas both S-IBP and S3R-IBP fail to capture the lowest values of the empirical diversity distribution (the vertical line for the S-IBP and S3R-IBP correspond to countries only having the bias term active defined in the Methods Section). Nevertheless, our model significantly outperforms S-IBP as it is able to capture the distribution for higher diversity values, i.e., S-IBP predicts a lower number of countries with high number of exports.

### 4.1 Quantitative Evaluation

We perform a quantitative evaluation of our model in terms of predictive accuracy, interpretability, strength and ability to capture the row marginal distribution of the input, using data from 2010. Simulations are run for 10 different train-test splits with a proportion of 90-10% entries. The burn-in period for the MCMC inference algorithm is 30,000 iterations, and results are averaged using the last 1,000 posterior samples. Table 5 compares our model against probabilistic matrix factorization (MF) [56], non-negative MF (NMF) [57], the standard Indian Buffet Process (IBP), and the sparse IBP (S-IBP) which uses $\alpha_B < 1$ in terms of mean and standard error of the test log-perplexities, and topic coherence.

**Table 5. Quantitative Evaluation of Accuracy and Interpretability.**

| 2010 SITC database ($C = 126$, $P = 744$) | | | | | |
|---|---|---|---|---|---|
| **Metric** | **MF** | **NMF** | **IBP** | **S-IBP** | **S3R-IBP** |
| Log Perplexity | $1.68 \pm 0.01$ | $1.61 \pm 0.01$ | $\mathbf{1.59 \pm 0.04}$ | $3.26 \pm 0.17$ | $1.62 \pm 0.01$ |
| Coherence | $-264.60 \pm 4.74$ | $-263.27 \pm 7.45$ | $-149.36 \pm 7.56$ | $-178.44 \pm 4.50$ | $\mathbf{-140.51 \pm 2.73}$ |
| 2010 HS database ($C = 123$, $P = 4890$) | | | | | |
| **Metric** | **MF** | **NMF** | **IBP** | **S-IBP** | **S3R-IBP** |
| Log Perplexity | $1.48 \pm 0.01$ | $\mathbf{1.47 \pm 0.01}$ | $1.58 \pm 0.01$ | $2.56 \pm 0.12$ | $1.57 \pm 0.02$ |
| Coherence | $-264.73 \pm 3.11$ | $-264.67 \pm 6.22$ | $-148.91 \pm 10.57$ | $-168.39 \pm 13.16$ | $\mathbf{-134.51 \pm 4.43}$ |

**Accuracy:** We use perplexity to measure predictive accuracy, i.e. the harmonic mean of the inverse test log likelihood. All the models present similar perplexity, except the S-IBP model, in which the sparseness restriction degrades its performance significantly. Our S3R-IBP has the same sparse restriction, but it has a more flexible prior that it is able to compensate the penalty in perplexity and perform close to to the non-sparse models, i.e. MF, NMF and IBP. The combination of the negative binomial and the stable-Beta process allows to match the perplexity performance of non sparse methods, but keeping the results interpretable, as we illustrate in the next paragraphs.

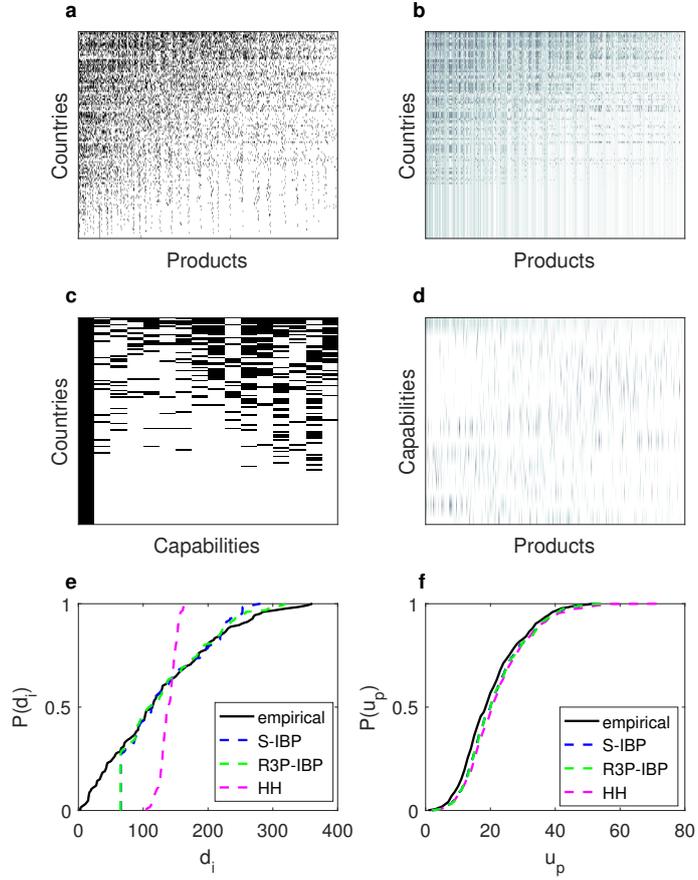

**Fig 8. Global properties generated by the model**: **a** Adjacency matrix for the empirical country-product matrix. **b** Adjacency matrix for the inferred country-product matrix, **c** Adjacency matrix for the inferred country-capability matrix. **d** Adjacency matrix for the inferred capability-product matrix. **d** Comparison of the fitted diversity cumulative distribution between the baseline, S-IBP and S3R-IBP. and the empirical country-product networks 2010. **f** Same as **e**, only for ubiquity. **a-d** Countries are ordered according to their diversity $d_c$, while products according to their ubiquity $u_p$. Darker shade indicates higher value.

**Interpretability:** We test the robustness of our results by performing comparison with with: i) results from Singular Value Decomposition (SVD) which serves as a baseline factorization method; ii) results corresponding to the year 1995; and iii) results estimated by using the Harmonized System (HS) rev. 1992 classification disaggregated to six digit level (see Section E of S1 Supporting Information for more details). The comparison with the SVD (provided in Table B in S1 Supporting Information), suggests that our approach significantly enhances interpretability of the latent factors in terms of both conciseness and precision of the clustering of products. Additionally, our findings are quite robust and stable in the sense that the evaluation of the model across different classifications and years does not produce any significant changes in the results, as detailed in Section E of S1 Supporting Information.

In order to assess semantic quality, we rely on the coherence [58], which is an often-used metric in topic modeling literature. The coherence $C_k$ of a feature is defined as

$$C_k = \sum_{m=2}^{M} \sum_{l=1}^{m-1} \log \frac{R(v_m^k, v_l^k) + 1}{R(v_l^k)} \qquad (4)$$

where $v_i^k$ is the $i$-th product with highest weight in factor $k$, and $M$ represents how many top products should be evaluated (here we take $M = 20$ products). Also $R(x)$ refers to the number of countries exporting product $x$, and $R(x, y)$ is the number of countries exporting both products $x$ and $y$. The closer coherence is to zero, the better.

The S3R-IBP outperforms not only NMF, but the IBP and S-IBP as well by far, as shown in Table 5, making it specially suitable for data exploration in high-dimensional count data scenarios. The non-sparse methods present a very low coherence, as expected. Intuitively, coherence measures the degree of homogeneity and agreement within a latent factor, and thus, will always degrade for non-sparse projections.

**Sparsity Structure:** We revisit the evaluation of the Baseline model, the S-IBP and the S3R-IBP model's fitting ability by comparing its Q-Q plot of the predicted diversification and ubiquity distribution versus the empirical distribution. The S3R-IBP and the S-IBP are by far superior than the Baseline model in fitting the distribution of diversity and ubiquity. Moreover, Fig. 9**a-b** reveals even more the differences between the S3R-IBP and S-IBP in their performance regarding the diversity distribution - it is clear that the S3R-IBP adequately captures the right tail of the distribution, whereas the S-IBP does not. Finally, both models exhibit promising results modeling the distribution of ubiquity (Fig. 9**d-e**). We estimate a two-sample Kolmogorov-Smirnov test for each predicted-real distribution combination as a means to statistically test the differences between them. The results are in line with the conclusions from the cumulative distribution and the Q-Q plot comparisons.

## 5 Discussion and Future Work

This work complements the studies in the field of economic complexity by introducing a probabilistic framework which leverages Bayesian non-parametric techniques. The framework aims at extracting the dominant features (capabilities) behind the comparative advantage in exported products. Based on economic evidence and trade data, we place a restricted Indian Buffet Process on the distribution of countries' capability endowment, appealing to a culinary metaphor to model the process of capability acquisition. We further extend the approach by introducing temporal dynamics in the acquisition of capabilities and introduction of new products in the countries' export portfolios. The overall approach comes with an adequate level of interpretability, as it produces a concise and economically plausible description of the instantiated capabilities.

The introduced framework offers a unifying qualitative and quantitative assessment of a country's productive structure. The factorization of the country-product matrix provides a descriptive view of the internal composition of each economy, and a direct way of comparison between different economies.

Several research directions remain for future work. First, it would be interesting to interpret the implications of the model from the perspective of evolutionary economic geography, and economic learning in general, for example with emphasize on the inter-industry and inter-regional learning channels. Second, we expect that the proposed framework would also be useful in the context of unrelated diversification in economic development (as in e.g. [59]) or, for example, in the study of industrial dynamics in the Varieties of Capitalism (VoC) framework [22].

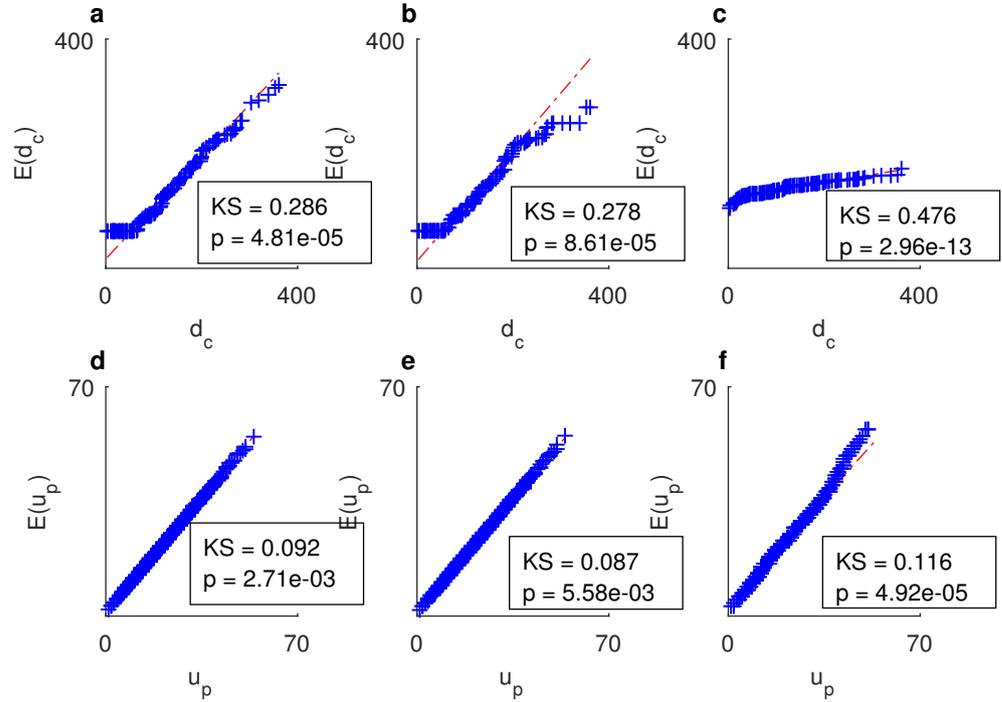

**Fig 9. Q-Q plots for the distribution inferred by the models**: **A** Diversity Q-Q plot for S3R-IBP. **B** Diversity Q-Q plot for the S-IBP. **C** Diversity Q-Q plot for the baseline model. **D** Ubiquity Q-Q plot for S3R-IBP. **E** Ubiquity Q-Q plot for S-IBP. **F** Ubiquity Q-Q plot for the baseline model.

Finally, there remains the question of how the predictions of the model may be put forward as policy recomendations. The first step in this direction is the extraction of relevant information than can be used by a particular country for developing industrial policies, for example by focusing on the easiest-to-be-acquired capabilities, or by following another country's development path in the capability space. In this context, our findings are only to be interpreted as the first step towards possible policy recommendations. The evaluation of the effects of specific policy recommendations, which may, for example, come in the form of subsidies or R&D incentives that the countries may provide to stimulate the production of specific products, is not covered here, but presents an important direction of future work.

# References

1. Lall S. Technological capabilities and industrialization. World development. 1992;20(2):165–186.

2. Lall S. The Technological structure and performance of developing country manufactured exports, 1985-98. Oxford development studies. 2000;28(3):337–369.

3. Hidalgo CA, Klinger B, Barabási AL, Hausmann R. The product space conditions the development of nations. Science. 2007;317(5837):482–487.

4. Hidalgo CA, Hausmann R. The building blocks of economic complexity. proceedings of the national academy of sciences. 2009;106(26):10570–10575.

5. Hausmann R, Hidalgo CA. The network structure of economic output. Journal of Economic Growth. 2011;16(4):309–342.

6. Hausmann R, Hidalgo CA, Bustos S, Coscia M, Simoes A, Yildirim MA. The atlas of economic complexity: Mapping paths to prosperity. Mit Press; 2014.

7. Hausmann R, Rodrik D. Economic development as self-discovery. Journal of development Economics. 2003;72(2):603–633.

8. Hausmann R, Hwang J, Rodrik D. What you export matters. Journal of economic growth. 2007;12(1):1–25.

9. Hidalgo C. Why information grows: The evolution of order, from atoms to economies. Basic Books; 2015.

10. Engerman SL, Sokoloff KL. Factor endowments, institutions, and differential paths of growth among new world economies. How Latin America Fell Behind. 1997; p. 260–304.

11. Felipe J, Kumar U, Abdon A, Bacate M. Product complexity and economic development. Structural Change and Economic Dynamics. 2012;23(1):36–68.

12. Rodrik D. What's so special about China's exports? China & World Economy. 2006;14(5):1–19.

13. Balassa B. The purchasing-power parity doctrine: a reappraisal. The Journal of Political Economy. 1964; p. 584–596.

14. Tacchella A, Cristelli M, Caldarelli G, Gabrielli A, Pietronero L. A new metrics for countries' fitness and products' complexity. Scientific reports. 2012;2.

15. Cristelli M, Gabrielli A, Tacchella A, Caldarelli G, Pietronero L. Measuring the intangibles: A metrics for the economic complexity of countries and products. PloS one. 2013;8(8):e70726.

16. Mariani MS, Vidmer A, Medo M, Zhang YC. Measuring economic complexity of countries and products: which metric to use? The European Physical Journal B. 2015;88(11):293.

17. Stojkoski V, Utkovski Z, Kocarev L. The impact of services on economic complexity: Service sophistication as route for economic growth. PloS one. 2016;11(8):e0161633.

18. Costanza R, Hart M, Talberth J, Posner S. Beyond GDP: The need for new measures of progress. The pardee papers. 2009;.

19. Fleurbaey M, Blanchet D. Beyond GDP: Measuring welfare and assessing sustainability. Oxford University Press; 2013.

20. Frenken K, Van Oort F, Verburg T. Related variety, unrelated variety and regional economic growth. Regional studies. 2007;41(5):685–697.

21. Neffke F, Henning M, Boschma R. How do regions diversify over time? Industry relatedness and the development of new growth paths in regions. Economic Geography. 2011;87(3):237–265.

22. Boschma R, Capone G. Institutions and diversification: Related versus unrelated diversification in a varieties of capitalism framework. Research Policy. 2015;44(10):1902–1914.

23. Jun B, Alshamsi A, Gao J, Hidalgo CA. Relatedness, knowledge diffusion, and the evolution of bilateral trade. arXiv preprint arXiv:170905392. 2017;.

24. Kogler DF, Rigby DL, Tucker I. Mapping knowledge space and technological relatedness in US cities. European Planning Studies. 2013;21(9):1374–1391.

25. Rigby DL. Technological relatedness and knowledge space: entry and exit of US cities from patent classes. Regional Studies. 2015;49(11):1922–1937.

26. Boschma R, Heimeriks G, Balland PA. Scientific knowledge dynamics and relatedness in biotech cities. Research Policy. 2014;43(1):107–114.

27. Petralia S, Balland PA, Morrison A. Climbing the ladder of technological development. Research Policy. 2017;46(5):956–969.

28. Kogut B, Zander U. Knowledge of the firm and the evolutionary theory of the multinational corporation. Journal of international business studies. 2003;34(6):516–529.

29. Fleming L, Sorenson O. Technology as a complex adaptive system: evidence from patent data. Research policy. 2001;30(7):1019–1039.

30. Balland PA, Rigby D. The geography of complex knowledge. Economic Geography. 2017;93(1):1–23.

31. Hartmann D, Guevara MR, Jara-Figueroa C, Aristarán M, Hidalgo CA. Linking economic complexity, institutions, and income inequality. World Development. 2017;93:75–93.

32. Arvis JF. How many dimensions do we trade in? Product space geometry and latent comparative advantage. 2013;.

33. Ghahramani Z. Probabilistic machine learning and artificial intelligence. Nature. 2015;521(7553):452.

34. Landauer TK, Foltz PW, Laham D. An introduction to latent semantic analysis. Discourse processes. 1998;25(2-3):259–284.

35. Escobar MD, West M. Bayesian Density Estimation and Inference Using Mixtures. Journal of the American Statistical Association. 1995;90(430):577–588. doi:10.1080/01621459.1995.10476550.

36. Griffiths TL, Ghahramani Z. The Indian Buffet Process: An Introduction and Review. Journal of Machine Learning Research. 2011;12:1185–1224.

37. Ruiz FJR, Valera I, Blanco C, Perez-Cruz F. Bayesian Nonparametric Comorbidity Analysis of Psychiatric Disorders. J Mach Learn Res. 2014;15(1):1215–1247.

38. Miller K, Jordan MI, Griffiths TL. Nonparametric Latent Feature Models for Link Prediction. In: Advances in Neural Information Processing Systems 22; 2009. p. 1276–1284.

39. Navarro DJ, Griffiths TL. Latent Features in Similarity Judgments: A Nonparametric Bayesian Approach. Neural Computation. 2008;20(11):2597–2628. doi:10.1162/neco.2008.04-07-504.

40. Chen M, Gao C, Zhao H. Phylogenetic Indian Buffet Process: Theory and Applications in Integrative Analysis of Cancer Genomics. arXiv:13078229 [math, q-bio, stat]. 2013;.

41. Pradier MF, Ruiz FJR, Perez-Cruz F. Prior Design for Dependent Dirichlet Processes: An Application to Marathon Modeling. PLOS ONE. 2016;11(1):e0147402. doi:10.1371/journal.pone.0147402.

42. Doshi-Velez F, Williamson SA. Restricted Indian Buffet Processes. arXiv:150806303. 2015;.

43. Boschma R. Relatedness as driver of regional diversification: A research agenda. Regional Studies. 2017;51(3):351–364.

44. Gopalan PK, Charlin L, Blei D. Content-based recommendations with Poisson factorization. In: Advances in Neural Information Processing Systems 27. Curran Associates, Inc.; 2014. p. 3176–3184.

45. Ricardo D. Principles of political economy and taxation. G. Bell and sons; 1891.

46. Teh YW, Gorur D. Indian Buffet Processes with Power-law Behavior. In: Advances in Neural Information Processing Systems 22. Curran Associates, Inc.; 2009. p. 1838–1846.

47. Pradier MF, Stojkoski V, Utkovski Z, Kocarev L, Perez-Cruz F. Indian Buffet Processes with Power-law Behavior. In: ICASSP proceedings; 2018.

48. Doshi-Velez F, Ghahramani Z. Correlated Non-parametric Latent Feature Models. In: Proceedings of the Twenty-Fifth Conference on Uncertainty in Artificial Intelligence. UAI '09. Arlington, Virginia, United States: AUAI Press; 2009. p. 143–150.

49. South A. rworldmap: A New R package for Mapping Global Data. R Journal. 2011;3(1).

50. Wickelmaier F. An introduction to MDS. Sound Quality Research Unit, Aalborg University, Denmark. 2003;46.

51. Jaffe AB, Trajtenberg M, Henderson R. Geographic localization of knowledge spillovers as evidenced by patent citations. the Quarterly journal of Economics. 1993;108(3):577–598.

52. Breschi S, Lissoni F. Mobility and social networks: Localised knowledge spillovers revisited. Università commerciale Luigi Bocconi; 2003.

53. Singh J. Collaborative networks as determinants of knowledge diffusion patterns. Management science. 2005;51(5):756–770.

54. Gao J, Jun B, Pentland A, Zhou T, Hidalgo CA, et al. Collective Learning in China's Regional Economic Development. arXiv preprint arXiv:170301369. 2017;.

55. Bruton HJ. Egypt's Development in the Seventies. Economic development and cultural change. 1983;31(4):679–704.

56. Mnih A, Salakhutdinov RR. Probabilistic matrix factorization. In: Advances in neural information processing systems; 2008. p. 1257–1264.

57. Schmidt MN, Winther O, Hansen LK. Bayesian non-negative matrix factorization. In: International Conference on Independent Component Analysis and Signal Separation. Springer; 2009. p. 540–547.

58. Newman D, Bonilla EV, Buntine W. Improving topic coherence with regularized topic models. In: Advances in neural information processing systems; 2011. p. 496–504.

59. Pinheiro FL, Alshamsi A, Hartmann D, Boschma R, Hidalgo C. Shooting Low or High: Do Countries Benefit from Entering Unrelated Activities? arXiv preprint arXiv:180105352. 2018;.

# Supporting Information

**S1 Supporting Information.** Additional description for the results presented in the manuscript:

1. Data Description
2. Indian Buffet Process and extensions
3. Sparse Three-parameter Restricted IBP
4. Inference and Settings
5. Robustness Results

S1 Supporting Information – Economic Complexity Unfolded: Interpretable Model for the Productive Structure of Economies

# A Data Description

## A.1 Trade Data

To empirically analyse the performance of our model we use the UN COMTRADE Standard International Trade Classification (SITC) rev.2 dataset, which disaggregates products to the four digit level, provided by the team of the Observatory of Economic Complexity. We focus our analysis on the year 2010. In order to clean unreliable or inadequately classified data, we restrict the dataset to the same countries that were used in the Atlas of Economic Complexity [5]. This leaves us with data on 126 countries and 744 products.

For the robustness check, we use the Harmonized System (HS) rev. 1992 classification disaggregated to six digit level (4890 products). The original data was collected by UN COMTRADE, and was further cleaned by the team of the Observatory of Economic Complexity (the HS data were also cleaned by the BACI team). They are available at http://atlas.media.mit.edu/en/resources/data/.

## A.2 Country-Product Network

Consider the graph $\mathcal{G} = (\mathcal{C}, \mathcal{P}, \mathcal{E})$ in which the vertices are partitioned into two disjoint sets: the set of countries $\mathcal{C}$ of cardinality $C$, and the set of products $\mathcal{P}$ exported by the countries, of cardinality $P$. An edge $(c, p)$ between a country $c \in [C]$ and a product $p \in [P]$ is present in the set of edges $\mathcal{E}$ if the country has a revealed comparative advantage in the export of that product:

$$M_{cp} = \begin{cases} 1, & \text{if } \text{RCA}_{cp} \geq 1 \\ 0, & \text{otherwise} \end{cases}. \tag{1}$$

In (1) $\text{RCA}_{cp}$ denotes Balassa's RCA index defined as:

$$\text{RCA}_{cp} = \frac{E_{cp}/\sum_p E_{cp}}{\sum_c E_{cp}/\sum_{c,p} E_{cp}}, \tag{2}$$

where $E_{cp}$ is the export of product $p$ by country $c$. $\text{RCA}_{cp} > 1$ indicates that country $c$'s share of product $p$ is larger than the product's share of the entire world market, thus "revealing" a comparative advantage of the country in the corresponding product.

The diversity (diversification) $d_c$ of a country $c$ is defined as the number of products in which it has comparative advantage, i.e. its degree in the network:

$$d_c = \sum_p M_{cp}, \tag{3}$$

Similarly, the ubiquity $u_p$ of a product is the number of countries that produce that product:

$$u_p = \sum_c M_{cp}. \tag{4}$$

# B Indian Buffet Process and extensions

## B.1 Indian Buffet Process

The Indian buffet process (IBP) is a stochastic process defining a probability distribution over equivalence classes of sparse binary matrices with a finite number of rows and unbounded number of columns [4]. Although the number of columns is potentially infinite, only a finite number of those will contain non-zero entries due to the finite nature of the observed data. Another important property of IBP-generated matrices is that they are exchangeable both in rows and columns, i.e., the order of the rows and columns are irrelevant. The IBP can be derived taking the limit as $K \to \infty$ of a finite binary matrix $\mathbf{Z} \in \{0,1\}^{C \times K}$, where $C$ is the number of observations, and $K$ is the number of latent features. Each element $Z_{ck}$ is distributed according to:

$$\begin{aligned} \pi_k &\sim \text{Beta}(\alpha/K, 1), \\ Z_{ck} &\sim \text{Bernoulli}(\pi_k), \end{aligned} \quad (5)$$

where $\pi_k$ is the probability of observing a non-zero value in column $k$ and $\mathbf{Z}_{c\bullet}$ is the $c$-th row representing sample $c$. We say that a feature $k$ is active for sample $c$ if $Z_{ck} = 1$. When $K \to \infty$, the above finite model tends to the IBP, denoted by:

$$\mathbf{Z} \sim \text{IBP}(\alpha), \quad (6)$$

where $\alpha$ is the mass parameter controlling the a priori activation probability of new features. Alternatively, the IBP can also be constructed based on its underlying De Finetti's representation, i.e., as a mixture of Bernoulli processes directed by a beta process[1]

$$\begin{aligned} \mu &\sim \text{BP}(1, \alpha, H), \quad (7)\\ \mathbf{Z}_{c\bullet} &\sim \text{BeP}(\mu), \quad (8) \end{aligned}$$

where $\mu$ is the directing measure, and $H$ is the probability base measure for the beta process [9].

In the IBP, the number of active features per row is distributed according to Poisson$(\alpha)$ and the total number of active features $K^+$, i.e., number of columns with non-zero entries, is distributed as Poisson$\left(\alpha \sum_{i=1}^{C} \left(\frac{1}{i}\right)\right)$. The single scalar parameter $\alpha$ has thus an effect on both sparsity density and sparsity structure of the latent matrix $\mathbf{Z}$. Such assumption might not always be appropriate, specially in data exploration applications where we expect a high number of latent features and a different sparsity degree per row in the latent matrix.

## B.2 Three-Parameter Indian Buffet Process

In the three-parameter IBP, the stick weights follow a more flexible distribution that cover power-law behaviors [8]. This can be achieved by replacing the beta process directing measure in the IBP by a stable-beta process. As its name indicates, this process can be fully specified by three parameters: $\alpha$ is the same mass parameter from the IBP that controls the *a priori* expected total number of non-zero entries in matrix $\mathbf{Z}$. Additionally, the stability exponent $\sigma \in [0,1)$ controls the power-law behavior of the model, and $\delta > -\sigma$ is the concentration parameter that affects the *a priori* number of ones per column. When $\delta = 1$ and $\sigma = 0$, we recover the standard IBP model.

Using the usual culinary metaphor of customers entering an Indian buffet restaurant and sequentially choosing dishes from an infinite buffet, the three-parameter IBP generalizes as follows:

- Customer 1 tries Poisson$(\alpha)$ number of dishes.

- Customer $c+1$ tries:
    - each dish with probability $\frac{m_k - \sigma}{c+\delta}$ for each dish that has previously been tried, where $m_k$ is the number of customers who previously sampled from dish $k$.
    - Poisson $\left(\alpha \frac{\Gamma(1+\delta)\Gamma(c+\delta+\sigma)}{\Gamma(c+1+\delta)\Gamma(\delta+\sigma)}\right)$ new dishes.

In such process, the number of hidden features is expected to grow as $O(C^\sigma)$. By introducing parameters $\delta$ and $\sigma$, matrix $\mathbf{Z}$ can have a more flexible sparsity structure, regardless of the sparsity degree which is controlled by $\alpha$. Compared to the Restricted IBP prior (described in the following), the 3-parameter IBP gives more flexibility regarding the feature weights, but has the disadvantage that the number of ones per-row is still a priori Poisson distributed, which might not be desirable in all situations, particularly in our analysis of international trade.

---

[1]Eq. 7 and 8 employ a common slight misuse of notation by ignoring the sticks position of the beta and Bernoulli processes.

## B.3 Restricted Indian Buffet Process

The Restricted IBP is a recently developed model that allows an arbitrary prior distribution to be placed over the number of active features underlying each observation [3]. A natural way to build such process is to replace the underlying Bernoulli processes in the IBP by *restricted* Bernoulli processes defined as:

$$\text{R-BeP}(\mathbf{Z}_{c\bullet}; \mu, g) = g(J_c) \cdot \frac{\prod_{k=1}^{\infty} \pi_k^{Z_{ck}} (1 - \pi_k^{1-Z_{ck}}) \mathbb{1}(\sum_K Z_{ck} = J_c)}{\sum_{Z' \in \mathcal{Z}} \prod_k \pi_k^{Z'_k} (1 - \pi_k)^{(1-Z'_k)} \mathbb{1}(\sum_K Z'_k = J_c)}, \tag{9}$$

where the directing measure $\mu = \sum_k \pi_k \delta_{\theta_k}$, $\pi_k$ and $\theta_k$ are the stick weight and location corresponding to each latent feature $k$, $\alpha$ is the same mass parameter present in the IBP case, and $g$ is the a priori distribution over the number of active features per sample. The Restricted IBP can thus be formulated as:

$$\mu \sim \text{BP}(1, \alpha, H), \tag{10}$$
$$\mathbf{Z}_{c\bullet} \sim \text{R-BeP}(\mu, g). \tag{11}$$

We thus have two degrees of freedom $\alpha$ and $g$ to control for sparsity degree and sparsity structure in matrix $\mathbf{Z}$. Note that columns are not exchangeable anymore, i.e., the parameter $g$ creates correlation among the features, which has to be dealt with during inference.

## C Sparse Three-parameter Restricted IBP

To decouple sparsity degree and sparsity structure in the latent matrix, we combine the advantages of both the Restricted IBP and three-parameter IBP into a single prior,

$$\mathbf{Z} \sim \text{3R-IBP}(\alpha, \delta, \sigma, g), \tag{12}$$

where the mass parameter $\alpha$ controls the sparsity degree of matrix $\mathbf{Z}$, $\delta$ is the concentration parameter that accounts for the degree of sharing between features, $\sigma$ is the stability exponent responsible for the power-law behavior of the stick weights, and $g$ is the a priori distribution over the number of ones per row.

Now, let $\mathbf{M} \in \mathbb{N}^{C \times P}$ be our input matrix of $C$ samples and $P$ dimensions. Using the three-parameter Restricted IBP prior, we build an infinite latent feature model for count data with Poisson likelihood and Gamma-distributed factors as follows

$$M_{cp} \sim \text{Poisson}(\mathbf{Z}_{c\bullet}\mathbf{B}_{\bullet p}), \tag{13}$$
$$B_{kp} \sim \text{Gamma}\left(\alpha_B, \frac{\mu_B}{\alpha_B}\right), \tag{14}$$
$$\mathbf{Z} \sim \text{3R-IBP}(\alpha, \delta, \sigma, g) \tag{15}$$

where $\alpha_B$ and $\mu_B$ are the shape and mean parameters of the prior Gamma distribution for each element of matrix $\mathbf{B}$. In this model, both matrices $\mathbf{Z}$ and $\mathbf{B}$ are non-negative and sparse, which makes the inferred latent variables very easy to interpret. In particular, sparsity in matrix $\mathbf{B}$ can be induced simply by choosing $\alpha_B \ll 1$.

In our particular application of international trade, we have $C$ countries, $P$ products and $K^+$ non-empty latent features to be inferred (which, in the spirit of economic complexity, we may refer to as capabilities). A given row $\mathbf{Z}_{c\bullet}$ captures which latent features are active for country $c$. On the other hand, matrix $\mathbf{B}$ represents the effect of each latent feature on every product. For instance, if a latent feature $k$ is active for a certain country, all products having high values in vector $\mathbf{B}_{k\bullet}$ will be more likely to be exported by that country.

To help further in the interpretation of features, we incorporate a *bias term* F0 into the model by forcing the first latent feature to be active for all countries[2]. By doing so, we are able to capture the average export probability for each product. Such approach has already been followed in [7] and [6] to alleviate identifiability problems in the inferred solution. This model can be seen as a probabilistic extension of non-negative matrix factorization where i) the number of latent features is not fixed a priori, ii) both matrices are sparse, and iii) soft-constraints on the expected latent sparsity structure are imposed through the prior.

---
[2]This feature may be seen as a basic capability present in every country.

# D  Inference and Settings

## D.1  Inference

Since exact computation of the posterior distribution for the latent variables is intractable, we resort to a Markov Chain Monte Carlo (MCMC) approach. Specifically, we use Gibbs sampling together with Metropolis-Hasting (MH). We utilize a finite-dimensional approximation for the latent measure $\pi$ by allowing at most $K$ features.

For each observation $M_{cp}$, we introduce the auxiliary variables $M'_{cp,1}, \ldots, M'_{cp,K}$ such that $M_{cp} = \sum_{k=1}^{K} M'_{cp,k}$, and $M'_{cp,k} \sim \text{Poisson}(Z_{ck} B_{kp})$ for $k = 1, \ldots, K$. Given such auxiliary variables, the model is conditionally conjugate, and a Gibbs sampler can be derived straightforwardly. The complete sampling algorithm is described in Algorithm 1.

---

**Algorithm 1** A single iteration of the MCMC inference procedure for the S3R-IBP model.

1: Sample each element of matrix $\mathbf{Z}$ using inclusion probabilities [1, 3].
2: Sample latent measure $\pi$ using MH steps [3].
3: Sample each element of $\mathbf{B}$ and $M'$ from their conditional distributions.
4: Sample hyperparameter $\alpha$ according to [2].

---

## D.2  Simulation Settings

We consider 10 different initializations for our model. For each simulation, there is a burn-in period for the MCMC inference algorithm of 10,000 iterations. After that, we average our results using 1,000 additional posterior samples. Regarding the parameters in our model, we choose $g = \text{Negative-Binomial}(r, q)$, with $r = [1, 2]$, and $q = [0.1, 0.3, 0.5]$. The results are equivalent using any of these priors. Here, we report the results for $r = 1$ and $q = 0.1$.

Additionally, we run experiments for each combination of $n = [1, 10, 20, 50]$ and $\sigma = [0, 0.25, 0.5, 0.75, 1]$. Even if the results did not vary considerably when changing those hyperparameters, setting $n > 1$ and $\sigma > 0$ allows for a higher *a priori* sparseness in the latent features and potential power-law behaviors in the stick weights respectively. All figures and tables correspond to $n = 50$, and $\sigma = 1$. The hyperparameters for the Gamma prior over $\alpha$ are shape and scale equal to one. Finally, $\alpha_B$ is set to 0.01 to induce sparsity, and $\mu_B$ is equal to 1.

# E  Robustness Results

We test the robustness of our results by comparing them with: i) results from Singular Value Decomposition (SVD) - a baseline factorization method; ii) results corresponding to the year 1995; and iii) results estimated by using the Harmonized System (HS) rev. 1992 classification disaggregated to six digit level. For an adequate comparison between all results, Table A presents a more complete representation than Figure 1 in the main text. For each capability $k$, we report the averaged number of countries $\bar{m}_k$ that have it, the top-5 products with highest weights $B_{kp}$ and a *representative country*, which we define as the country that has the least number of active capabilities among those that possess capability $k$. We also report the average number of active capabilities $\bar{J}_c$ for each representative country $c$.

## E.1  Comparison with SVD

In Table B we report the Top-5 products with sorted highest weights from the Top-15 features learned via SVD. By comparing SVD to S3R-IBP, it is evident that our model is able to give much shorter and concise descriptions, as weights decrease at a faster pace and the largest weights are considerably larger. Moreover, the products listed in each feature of SVD come from a mixture of several different production elements (e.g. F2 includes musical accessories, friction materials, vegetables, meat and even metal products), whereas the S3R-IBP list is much more homogeneous. Thus, it is easy to conclude that our approach enhances interpretability of the latent factors in terms of both conciseness and precision, when compared to a baseline matrix factorization method.

## E.2 Comparison across years

Table C presents the results for the year 1995 estimated with our model. The results are very similar as the ones presented in Table A. For instance, feature F4 that was learned in 2010 corresponds to feature F3 learned in 1995, whereas feature F11 corresponds to the same products (e.g., vehicles) in both years. Additionally, features that are present in most countries (except the bias) appear in both years. Although labels might be switched, which is to be expected in these latent feature models, we infer similar capabilities for 1995 and 2010.

## E.3 Comparison across datasets

Table D displays the results for 2010 estimated for the HS classification dataset ($C = 123$, $P = 4890$). Since the number of products in this dataset is much higher than for the SITC classification, our model infers a higher number of capabilities. Indeed, the HS classification offers more information about the productive structure of economies. Despite the difference in the total number of capabilities, there is a striking resemblance among the capabilities in both cases. For example, we can clearly associate capability F2 from Table A to capability F4 from D, or the vehicle feature F11, with feature F14. Interestingly, the HS classification offers more granular capabilities to describe a single capability in the SITC case, e.g., F1, F2 and F4 in Table D corresponds to F2 in Table A.

**S1 Table A.** Complete List of Capabilities found by the S3R-IBP model in 2010 through the SITC classification.

| Id | $\bar{m}_k$ | Top-5 products with sorted highest weights ($B_{kp}$) associated | Repr. countries ($\bar{J}_c$) |
|---|---|---|---|
| F0 | 126 | Non-Coniferous Worked Wood (0.40), Bran and Other Cereals Residues (0.39), Miscellaneous Non-Iron Waste (0.38), Unwrought Lead (0.38), Bones, Ivory and Horns (0.37) | - |
| F1 | 38.67 | Vegetables (0.60), Fruit or Vegetable Juices (0.54), Miscellaneous Fruit (0.50), Frozen Vegetables (0.48), Apples (0.47) | Peru (2.00) |
| F2 | 46.11 | Synthetic Knitted Undergarments (0.76), Miscellaneous Feminine Outerwear (0.74), Miscellaneous Knitted Outerwear (0.73), Men's Shirts (0.70), Blouses (0.67) | Sri Lanka (2.00) |
| F3 | 18.27 | Miscellaneous Animal Oils (0.78), Bovine and Equine Entrails (0.72), Bovine meat (0.68), Preserved Milk (0.63), Equine (0.62) | Paraguay (2.00) |
| F4 | 21.39 | Synthetic Woven Fabrics (0.74), Non-retail Synthetic Yarn (0.60), Woven Fabric of less than 85% Discontinuous Synthetic Fibres (0.60), Woven Fabrics of More Than 85% Discontinuous Synthetic Fiber (0.58), Yarn of Less Than 85% Synthetic Fibers (0.53) | United Arab Emirates (2.82) |
| F5 | 16.53 | Miscellaneous Electrical Machinery (0.76), Vehicles Stereos (0.72), Miscellaneous Data Processing Equipment (0.64), Video and Sound Recorders (0.57), Calculating Machines (0.55) | Malaysia (3.00) |
| F6 | 45.93 | Baked Goods (0.67), Metal Containers (0.62), Miscellaneous Edibles (0.59), Miscellaneous Articles of Paper (0.59), Miscellaneous Organic Surfactants (0.58) | Costa Rica (2.06) |
| F7 | 21.95 | Measuring Controlling Instruments (0.61), Mathematical Calculation Instruments (0.59), Miscellaneous Electrical Instruments (0.57), Miscellaneous Heating and Cooling Equipment (0.51), Parts of Office Machines (0.49) | Malaysia (3.00) |
| F8 | 33.23 | Miscellaneous Articles of Iron (0.65), Carpentry Wood (0.61), Miscellaneous Manufactured Wood Articles (0.60), Sawn Wood Less Than 5mm Thick (0.56), Electric Current (0.51) | Russia (2.93) |
| F9 | 32.12 | Miscellaneous Rotating Electric Plant Parts (0.66), Control Instruments of Gas or Liquid (0.58), Valves (0.57), Miscellaneous Rubber (0.56), Miscellaneous Articles of Plastic (0.55) | Philippines (4.01) |
| F10 | 33.00 | Improved Wood (0.71), Mineral Wool (0.62), Central Heating Equipment (0.62), Aluminium Structures (0.62), Harvesting Machines (0.60) | Belarus (4.20) |
| F11 | 31.14 | Vehicles Parts and Accessories (0.59), Cars (0.58), Iron Wire (0.53), Trucks and Vans (0.53), Air Pumps and Compressors (0.50) | Belarus (4.20) |
| F12 | 11.04 | Synthetic Rubber (0.87), Acrylic Polymers (0.85), Silicones (0.76), Miscellaneous Polymerization Products (0.71), Tinned Sheets (0.65) | North Korea (3.99) |
| F13 | 18.67 | Aldehyde, Ketone and Quinone-Function Compounds (0.68), Glycosides and Vaccines (0.67), Medicaments (0.65), Inorganic Esters (0.64), Cyclic Alcohols (0.62) | Ireland (4.34) |
| F14 | 14.87 | Parts of Metalworking Machine Tools (0.74), Interchangeable Tool Parts (0.72), Polishing Stones (0.69), Tool Holders (0.66), Miscellaneous Metalworking Machine-Tools (0.54) | Israel (5.97) |
| F15 | 23.29 | Miscellaneous Pumps (0.51), Ash and Residues (0.45), Chemical Wood Pulp of sulphite (0.44), Rolls of Paper (0.43), Worked Nickel (0.43) | Russia (2.93) |

From left to right, $\bar{m}_k$ is the averaged number of countries having latent feature $k$ active, we list the top-5 products with highest weights $B_{kp}$; a *representative country* is the country that has the least number of capabilities among those possessing feature $k$. $\bar{J}_c$ is the averaged number of active features for each representative country $c$.

**S1 Table B.** Top-15 latent features inferred using the Singular Value Decomposition through the SITC classification.

| Id | Top-5 products with sorted highest weights ($B_{kp}$) associated |
|---|---|
| F1 | Miscellaneous Non-Ferrous Ores (0.40), Petroleum Gases (0.40), Miscellaneous Textile Articles (0.37), Zinc Ore (0.32), Miscellaneous Bituminous Mixtures (0.31) |
| F2 | Sound Recording Media (0.38), Asbestos Products (0.38), Potatoes (0.37), Silver (0.35), Pig Meat (0.32) |
| F3 | Thin Iron Sheets (0.42), Miscellaneous Food-Processing Machinery (0.41), Baked Goods (0.41), Miscellaneous Animal Entrails (0.34), Basketwork (0.34) |
| F4 | Perfumery and Cosmetics (0.45), Miscellaneous Gas Turbines (0.38), Cut Paper (0.35), Miscellaneous Cereal Grains (0.33), Herbicides (0.32) |
| F5 | Bovine (0.49), Miscellaneous Refrigeration Equipment (0.43), Radioactive Chemicals (0.41), Blocks of Iron and Steel (0.41), Rape Seeds (0.40) |
| F6 | Wheat Flour (0.34), Iron and Steel Forging (0.29), Printing Ink (0.29), Waste Paper (0.28), Aluminum (0.26) |
| F7 | Miscellaneous Oil Seeds and Fruits (0.47), Bones, Ivory and Horns (0.44), Temporarily Preserved Fruit (0.43), Cotton Seed Oil (0.42), Inorganic Bases (0.39) |
| F8 | Prepared Explosives (0.48), Confectionary Sugar (0.39), Cigarretes (0.38), Coke (0.37), Miscellaneous Hides and Skins (0.34) |
| F9 | Fish, preserved (0.44), Fresh Fish (0.43), Miscellaneous Animal Origin Materials (0.40), Oranges (0.37), Sheep and Goat Meat (0.37) |
| F10 | Wood and Animal Hair Waste (0.46), Miscellaneous Carpets (0.42), Wool Carpets (0.41), Wool Yarn (0.40), Degreased Sheep Wool (0.38) |
| F11 | Tin (0.41), Vehicles Stereos (0.40), Copper (0.36), Miscellaneous Articles of Paper (0.36), Petroleum Gases (0.36) |
| F12 | Gypsum and Other Calcareous Stone (0.42), Sausage (0.34), Special Products of Textile (0.32), Movie Cameras and Equipment (0.30), Iron Shapes (0.29) |
| F13 | Cigarretes (0.50), Worked Tin and Alloys (0.43), Aluminum (0.38), Bicycles (0.38), Raw Sheep Skin without Wool (0.38) |
| F14 | Precious Metal Ores (0.50), Gold (0.48), Diamonds (0.47), Unmounted Precious Stones (0.43), Electrical Transformers (0.38) |
| F15 | Sulphur (0.40), Fuel Wood and Charcoal (0.34), Miscellaneous Unmilled Cereals (0.33), Household Refrigeration (0.33), Decorative Wood (0.33) |

**S1 Table C.** Complete List of Capabilities found by the S3R-IBP model in 1995 through the SITC classification.

| Id | $\bar{m}_k$ | Top-5 products with sorted highest weights ($B_{kp}$) associated | Repr. countries ($\bar{J}_c$) |
|---|---|---|---|
| F0 | 125 | Miscellaneous Non-Iron Waste (0.45), Men's Shirts (0.39), Miscellaneous Live Animals (0.39), Lubricating Petroleum Oils (0.38), Raw Goat Skins (0.37) | - |
| F1 | 45.46 | Fruit Jams (0.51), Baked Goods (0.49), Confectionary Sugar (0.49), Miscellaneous Metal Articles (0.47), Miscellaneous Beverages (0.46) | Jordan (2) |
| F2 | 51.18 | Womens Knitted Outerwear (0.81), Womens Coats (0.8), Skirts (0.78), Mens Jackets (0.77), Miscellaneous Feminine Outerwear (0.76) | Bangladesh (2) |
| F3 | 15.22 | Woven Fabric of less than 85% Discontinuous Synthetic Fibres (0.75), Synthetic Woven Fabrics (0.68), Embroidery (0.63), Fabrics of more than 85% discontinuous regenerated fibres (0.63), Miscellaneous Woven Fabrics (0.63) | Pakistan (3) |
| F4 | 13.86 | Miscellaneous Electrical Machinery (0.81), Computer Peripherals (0.79), Computer Parts and Accessories (0.79), Electrical Resistors (0.68), Diodes, Transistors and Photocells (0.67) | Singapore (3) |
| F5 | 10.04 | Cameras (0.93), Calculating Machines (0.83), Video and Sound Recorders (0.81), Vehicles Stereos (0.8), Recorded Audio Players (0.75) | Singapore (3) |
| F6 | 19.84 | Miscellaneous Articles of Plastic (0.62), Mirrors (0.61), Plastic Lamps (0.59), Miscellaneous Non-Electrical Machinery Parts (0.53), Locksmith Hardware (0.5) | Hong Kong (6) |
| F7 | 43.7 | Fiberboard (0.69), Miscellaneous Furniture (0.48), Chemical Wood Pulp of sulphite (0.46), Tissue Paper (0.46), Wood Boxes (0.45) | Zimbabwe (3.02) |
| F8 | 11.1 | Organic Chemical Products (0.65), Phenols(0.58), Gas, Liquid and Electric Meters (0.57), Chemical Products (0.53), Fungicides (0.5) | Israel (6.98) |
| F9 | 40.04 | Iron Coils (0.59), Thin Iron Sheets (0.54), Thick Iron Sheets (0.52), Metal Cables (0.51), Iron Wire (0.47) | Jordan (3) |
| F10 | 15.04 | Valves (0.89), Industrial Furnaces and Ovens (0.7), Centrifugal Pumps (0.68), Miscellaneous Refrigeration Equipment (0.64), Pulley System Parts (0.64) | Bosnia and Herzegovina (4.02) |
| F11 | 15.14 | Vehicles Parts and Accessories (0.71), Motor Vehicles Piston Engines (0.66), Miscellaneous Refractory Goods (0.63), Miscellaneous Rubber (0.58), Piston Engine Parts (0.53) | Mexico (6.06) |
| F12 | 22 | Miscellaneous Mineral Materials (0.78), Mineral Wool (0.73), Varnishes and Lacquers (0.64), Glass Fiber Fabrics (0.64), Miscellaneous Printed Matter | Norway (4) |
| F13 | 12.06 | Amine-Function Compounds (0.88), Inorganic Esters (0.73), Oxygen-Function Amino-Compounds (0.66), Organo-Sulphur Compounds (0.65), Glues (0.63) | Israel (6.98) |
| F14 | 23.04 | Iron Tubes (0.83), Miscellaneous Machinery (0.64), Fasteners (0.60), Miscellaneous Parts of Lifting Machinery (0.59), Miscellaneous Iron Tubes and Pipes (0.55) | Georgia (2) |
| F15 | 14.74 | Interchangeable Tool Parts (0.74), Miscellaneous Pump Parts (0.67), Miscellaneous Liquid Pump Parts (0.66), Miscellaneous Printing Machines (0.64), Miscellaneous Agricultural Machinery (0.62) | Israel (6.98) |
| F16 | 23.26 | Iron Ore Agglomerates (0.56), Ferro-alloys (0.53), Prepared Explosives (0.53), Iron Ore (0.49), Inorganic Bases (0.47) | Venezuela (2.12) |

From left to right, $\bar{m}_k$ is the averaged number of countries having latent feature $k$ active, we list the top-5 products with highest weights $B_{kd}$; a *representative country* is the country that has the least number of capabilities among those possessing feature $k$. $\bar{J}_c$ is the averaged number of active features for each representative country $n$.

**S1 Table D.** Complete List of Capabilities found by the S3R-IBP model in 2010 through the HS classification.

| Id | $\bar{m}_k$ | Top-5 products with sorted highest weights ($B_{kp}$) associated | Repr. countries ($\bar{J}_c$) |
|---|---|---|---|
| F0 | 123 | Waste or scrap, aluminium (0.39), Wheat bran, sharps, other residues (0.39), Lumber, non-coniferous nes (0.37), Oils petroleum, bituminous, distillates, except crude (0.37), Lead unwrought containing mostly antimony (0.37) | - |
| F1 | 44 | Mens, boys trousers & shorts, material nes, not knit (0.72), Womens, girls blouses & shirts, material nes, not knit (0.72), Mens, boys jackets & blazers, material nes, not knit (0.72), Womens, girls skirts, of material nes, not knit (0.71), Womens, girls blouses & shirts, of material nes, knit (0.69) | Ethiopia (2) |
| F2 | 40.26 | Womens, girls suits, of wool or hair, not knit (0.64), Womens, girls dresses, of wool or hair, knit (0.56), Womens, girls overcoats, etc, of wool or hair, knit (0.55), Mens, boys suits, of materials nes, knit (0.52), Womens, girls garments nes, of wool or hair, not knit (0.51) | Australia (3) |
| F3 | 38.77 | Ammonium nitrate limestone etc mixes, pack >10 kg (0.54), Plastic builders ware nes (0.53), Scarifiers, cultivators, weeders and hoes (0.52), Waste or scrap, of stainless steel (0.51), Furniture parts nes (0.51) | Albania (4) |
| F4 | 37.98 | Womens, girls blouses & shirts, of cotton, knit (0.8), Mens, boys shirts, of cotton, knit (0.75), Pullovers, cardigans etc of cotton, knit (0.75), Womens, girls trousers & shorts, of cotton, knit (0.73), Mens, boys shirts, of manmade fibres, knit (0.71) | Ethiopia (2) |
| F5 | 34 | Parts for electric motors and generators (0.62), Electric heating resistors (0.59), Parts of cycle & vehicle light, signal, etc equipment (0.59), Parts of electrical transformers and inductors (0.59), Lock parts, etc, of base metal (0.56) | Moldova (5) |
| F6 | 31.31 | Polymer based paints & varnishes nes, aqueous medium (0.46), Synthetic organic products used as luminophores (0.43), Chain and parts thereof of copper (0.41), Insulated winding wire, nes (0.38), Boxes, moulding, for metal foundry (0.36) | Albania (4) |
| F7 | 30 | Bars, rods and other profiles, aluminium alloyed (0.64), Plastic doors and windows and frames thereof (0.64), Wooden cases, boxes, crates, drums and containers (0.63), Angles/shapes/sections, iron or non-alloy steel, nes (0.61), Sheet etc, cellular of polymers of styrene (0.60) | Moldova (5) |
| F8 | 28 | Springs, iron or steel, except helical/leaf (0.59), Foil, aluminium, not backed, rolled but nfw, <0.2mm (0.52), Chain, iron or steel, nes (0.50), Railway passenger and special purpose coaches (0.45), Parts, accessories nes, metal cutting machine tools (0.43) | Norway (3.43) |
| F9 | 26.88 | Trailer/non-mechanically propelled vehicle parts nes (0.65), Milk and cream not concentrated nor sweetened <6% fa (0.65), Poultry, domestic, whole, fresh or chilled (0.65), Sausages, similar products of meat, meat offal & bloo (0.64), Milk not concentrated nor sweetened 1-6% fat (0.63), | Ireland (4.34) |
| F10 | 25 | Drinking glasses, except lead crystal or glass ceramic (0.61), Refractory bricks etc >50% alumina or silica (0.55), Wheels including parts/accessories for motor vehicles (0.54), Woven fabric polyester + wool or hair, nes (0.52), Glass mirrors, unframed (0.52) | Australia (3) |
| F11 | 24 | Textile products and articles for technical uses, nes (0.59), Parts, laboratory/industrial heating/cooling machinery (0.57), Medical, surgical or laboratory sterilizers (0.54), Electrical relays for 60 - 1,000 volts (0.54), Electrodes etc of base metal or metal carbide, nes (0.51) | Canada (8) |
| F12 | 24 | Parts agricultural, forestry, bee-keeping machines nes (0.78), Parts of agricultural machinery (0.62), Parts for soil preparation or cultivation machinery (0.60), Rye (0.59), Hydraulic power engines/motors, linear acting (0.57) | Norway (3.43) |

**S1 Table D Continued.**

| Id | $\bar{m}_k$ | Top-5 products with sorted highest weights ($B_{kp}$) associated | Repr. countries ($\bar{J}_c$) |
|---|---|---|---|
| F13 | 23 | Cranes or derricks nes (0.66), Bars, rods & profiles of copper-zinc base alloys (0.62), Polishes, creams etc. for maintenance of woodwork (0.54), Num controlled machine tools to bend, fold, etc, meta (0.54), Acrylic & vinyl polymer based paint, varnish, in water (0.51) | South Africa (7.39) |
| F14 | 22 | Motor vehicle parts nes (0.77), Parts and accessories of bodies nes for motor vehicle (0.68), Rubber tube, pipe, hose textile-reinforced no fitting (0.67), Medium Diesel Engine Cars (0.65), Motor vehicle mountings, fittings, of base metal, nes (0.64) | Canada (8) |
| F15 | 22 | Embroidery of natural textile fibres except cotton (0.67), Woven fabric>85% synth nes+cotton, >170g/m2 unbl/blch (0.64), Woven fabric>85% synth nes + cotton,<170g/m2 yarn dye (0.63), Unglazed ceramic mosaic tiles etc, <7cm wide (0.60), Carpets of yarn nes, woven pile, not made up, nes (0.55) | Albania (4) |
| F16 | 17.91 | Grinding/polishing machines for stone, ceramics, glass (0.55), Dividing heads/attachments nes for machine tools (0.53),Tools for milling (0.53), Industrial electric resistance heated furnaces & oven (0.52), Blades for kitchen appliances & food industry machine (0.51) | Estonia (11) |
| F17 | 17 | Sheet/film not cellular/reinf amino-resins (0.57), Anti-oxidisers and stabilizers for rubber or plastics (0.51), Nitrile-function compounds, nes (0.42), Esters of inorganic acids, nes, their salts, derivs (0.41), Parts and accessories for drawing, etc instruments (0.41) | Canada (8) |
| F18 | 16.77 | Parts of milking machines and dairy machinery (0.67), Eels, fresh or chilled, whole (0.63), Industrial machinery nes for food, drink preparation (0.61), Cod, fresh or chilled, whole (0.59), Chemical industry products, preparations, mixtures (0.53) | Moldova (5) |
| F19 | 16 | Gasket sets, other joints of similar composition (0.65), Lubricating preparations, zero petroleum content, nes (0.60), Spectrometers, spectrophotometers, etc using light (0.57), Staple fibres of nylon, polyamides, not carded, combe (0.52), Organo-sulphur compounds, nes (0.51) | Singapore (9) |
| F20 | 15.4 | Synthetic organic tanning substances (0.69), Citrus fruits, otherwise prepared or preserved (0.54), Other finishing agents of a kind used in the leather or like industries (0.53), Inorganic tanning and pre-tanning preparations (0.53), Citrus juice nes (one fruit) not fermented or spirite (0.5) | Australia (3) |
| F21 | 15.27 | Heterocyclic compound,oxygen hetero-atom(s) only,nes (0.73), Heterocyclic compounds with unfused pyridine ring, nes (0.71), Lactones, other than coumarins (0.59), Printing ink, other than black (0.59), Heterocyclic compds with an unfused triazine ring nes (0.58) | Ireland (4.34) |
| F22 | 15 | Tungsten unwrought, bars/rods simply sintered, scrap (0.63), Mica plates, sheets and strips (0.6),Parts and accessories for flashlights and flashbulbs (0.55), Circular saw blades, working part other than steel (0.52), Base metals clad with silver, semi-manufactured (0.49) | Estonia (11) |
| F23 | 14 | Horses, live except pure-bred breeding (0.82), Bovine cuts boneless, fresh or chilled (0.71), Cheese, grated or powdered, of all kinds (0.70), Bovine livers, frozen (0.7), Seed, rye grass, for sowing (0.70) | Australia (3) |
| F24 | 14 | Olives, prepared or preserved, not frozen/vinegar (0.75), Olive oil, fractions, blends, not chemically modified (0.70), Olive oil, fractions, refined, not chemically modifie (0.70), Tomatoes nes, prepared or preserved, not in vinegar (0.70), Olive oil, virgin(0.68) | Belarus (6) |

**S1 Table D Continued.**

| Id | $\bar{m}_k$ | Top-5 products with sorted highest weights ($B_{kp}$) associated | Repr. countries ($\bar{J}_c$) |
|---|---|---|---|
| F25 | 13 | Woven cotton nes, <85% +manmade fibre, <200g/m2 dyed (0.8), Woven fabric synthetic filament <85% +cotton, dyed ne (0.79), Plain weave cotton, >85% 100-200g/m2, dyed (0.75), Woven cotton nes, >85% <200g/m2, printed (0.71), Woven cotton nes, >85% >200g/m2, dyed, nes (0.69) | Moldova (5) |
| F26 | 13 | Belts and bandoliers of leather or composition leather (0.84), Woven fabric synthetic staple fibre with manmade, nes (0.71), Sheep or lamb skin leather, nes (0.68), Twill cotton except denim, >85% >200g/m2, yarn dyed (0.63), Woven fabric <85% artificial staple+manmade fibre dye (0.59) | Belarus (6) |
| F27 | 12 | Cotton yarn <85% single uncombed >714dtex, not retail (0.85), Cotton yarn <85% multiple combed <125 dtex, not retai (0.75), Cotton yarn <85% single combed >714 dtex, not retail (0.71), Cotton yarn <85% single combed 192-125 dtex,not retai (0.70), Cotton yarn <85% single combed <125 dtex, not retail (0.70) | Canada (8) |
| F28 | 11.04 | Adipic acid, its salts & esters (0.64), Diols except ethylene and propylene glycol (0.62), Coated rods/cored wire for flame solder/braze/weld (0.61), Film in rolls, width 105-610 mm nes (0.60), Acrylic acid esters (0.59) | Belarus (6) |
| F29 | 10 | Parts and accessories of recorders except cartridges (1.02), Cast, drawn or float glass sheet, edge worked or bent (0.84), Soya sauce (0.84), Electri capacitors, fixed, ceramic, single layer (0.83), Permanent magnets & articles intended as magnets, nes (0.79) | Moldova (5) |
| F30 | 9 | 1-chloro-2,3-epoxypropane(epichlorohy-drin) (0.76), Dichloromethane (methylene chloride) (0.74), Halides & halide oxides of non-metals (not chlorides) (0.64), Chloroform (trichloromethane) (0.64), Cresols, salts (0.61) | Albania (4) |
| F31 | 6 | Bituminous coal, not agglomerated (0.81), Balls, iron/steel, forged/stamped for grinding mills (0.77), Iron ore, concentrate, not iron pyrites, agglomerated (0.61), Residual lyes from the manufacture of wood pulp (0.60), Palladium in semi-manufactured forms (0.56) | Australia (3) |
| F32 | 6 | Vinyl polymers, halogenated olefins, primary form, ne (0.89), Polyvinyl alcohols in primary form (0.64), Methyloxirane (propylene oxide)(0.63), Ultra-violet or infra-red lamps, arc lamps (0.57), Tin alloys unwrought (0.56) | Singapore (9) |
| F33 | 6 | Sound reproducing apparatus, non-recording (0.85), Radio receivers, portable, with sound reproduce/recor (0.80), Erasers (vulcanised rubber) (0.76), Multiple loudspeakers, mounted in single enclosure (0.76), Telephone sets (0.70) | Canada (8) |
| F34 | 5 | Paper, filter, cut to size or shape (0.77), Carbon tetrachloride (0.61), Cellulose acetates, non-plasticised, in primary forms (0.54), Polyvinyl chloride nes, not plasticised, primary form (0.50), Spacecraft, satellites and spacecraft launch vehicles (0.49) | United Kingdom (19) |
| F35 | 4 | Articles for Christmas festivities (1.05), Parts nes, for dolls representing only human beings (0.96), Dolls representing only human beings (0.91), Lighting sets of a kind used for Christmas trees (0.89), Stuffed toys - animals or non-human creatures (0.88), | Moldova (5) |
| F36 | 3 | Antimony, articles thereof, waste or scrap (0.91), Electric toasters, domestic (0.91), Manganese, articles thereof, waste or scrap (0.79), Rare-earth metals, scandium and yttrium (0.68), Fabric impregnated, coated, covered with polyurethane (0.66) | Singapore (9) |
| F37 | 3 | Molybdenum concentrates, roasted (0.73), Molybdenum ores and concentrates except roasted (0.70), Cameras for 35 mm roll film except single lens reflex (0.63), Lettuce, fresh or chilled except cabbage lettuce (0.61), Propionic acid, its salts & esters (0.55), | Canada (8) |

**S1 Table D Continued.**

| Id | $\bar{m}_k$ | **Top-5 products with sorted highest weights ($B_{kp}$) associated** | Repr. countries ($\bar{J}_c$) |
|---|---|---|---|
| F38 | 2 | Chlorobenzene, o-dichlorobenzene and p-dichlorobenzen (0.59), Cresols, salts (0.53), Aminohydroxynaphthalenesulphonic acids and salts (0.5), Diphenylamine, derivatives, salts thereof (0.49), Zinc sulphide (0.48) | India (18) |
| F39 | 1 | Parts of cathode-ray tubes (0.51), Moulds for rubber or plastic, nes (0.50), Bolts/screws nes, with/without nut/washer, iron/steel (0.46), Plates, sheet, strip and foil, nickel, not alloyed (0.45), Cotton yarn <85% single uncombed >714dtex, not retail (0.45), | Moldova (5) |
| F40 | 1 | Parts and accessories for photo-copying apparatus (0.52), Wagon handling equipment (0.47), Thread rolling machines for working metal, etc (0.46), Adipic acid, its salts & esters (0.46), Xylenols, salts (0.44) | Canada (8) |
| F41 | 1 | Powders, alloy steel (0.46), Injection-moulding machines for rubber or plastic (0.43), Chem wood pulp, sulphite, non-coniferous, unbleached (0.43), Apparatus for electro-plating, electrolysis, etc (0.42), Spinning spindles, spindle flyers, spinning rings (0.41) | Canada (8) |
| F42 | 1 | Seed, rye grass, for sowing (0.44), Pneumatic hand tool parts (0.41), Bed linen, of material nes (0.41), Bird skins and feathers, articles therefrom (0.39), Cartridges for rivet etc tools, humane killers, etc(0.38) | South Africa (7.39) |

From left to right, $\bar{m}_k$ is the averaged number of countries having latent feature $k$ active, we list the top-5 products with highest weights $B_{kd}$; a *representative country* is the country that has the least number of capabilities among those possessing feature $k$. $\bar{J}_c$ is the averaged number of active features for each representative country $n$.

# Bibliography

[1] Nibia Aires. "Algorithms to find exact inclusion probabilities for conditional Poisson sampling and Pareto sampling designs". In: *Methodology and Computing in Applied Probability* 1.4 (1999), pp. 457–469.

[2] Cedric Archambeau, Balaji Lakshminarayanan, and Guillaume Bouchard. "Latent IBP compound Dirichlet allocation". In: *IEEE transactions on pattern analysis and machine intelligence* 37.2 (2015), pp. 321–333.

[3] Finale Doshi-Velez and Sinead A. Williamson. "Restricted Indian Buffet Processes". In: *arXiv preprint arXiv:1508.06303* (2015).

[4] Thomas L. Griffiths and Zoubin Ghahramani. "The Indian Buffet Process: An Introduction and Review." In: *Journal of Machine Learning Research* 12 (2011), pp. 1185–1224.

[5] César A Hidalgo et al. "The product space conditions the development of nations". In: *Science* 317.5837 (2007), pp. 482–487.

[6] Melanie F. Pradier, Francisco J. R. Ruiz, and Fernando Perez-Cruz. "Prior Design for Dependent Dirichlet Processes: An Application to Marathon Modeling". In: *PLOS ONE* 11.1 (Jan. 2016), e0147402. ISSN: 1932-6203. DOI: `10.1371/journal.pone.0147402`.

[7] Francisco J. R. Ruiz et al. "Bayesian Nonparametric Comorbidity Analysis of Psychiatric Disorders". In: *J. Mach. Learn. Res.* 15.1 (Jan. 2014), pp. 1215–1247. ISSN: 1532-4435.

[8] Yee W. Teh and Dilan Gorur. "Indian Buffet Processes with Power-law Behavior". In: *Advances in Neural Information Processing Systems 22*. Ed. by Y. Bengio et al. Curran Associates, Inc., 2009, pp. 1838–1846.

[9] Romain Thibaux and Michael I. Jordan. "Hierarchical beta processes and the Indian buffet process". In: *International Conference on Artificial Intelligence and Statistics*. 2007, pp. 564–571.